%% file: iplong.tex
 \documentclass[twocolumn,showpacs,amsmath,prd,amssymb,showkeys,superscriptaddress,aps,longbibliography,nofootinbib]{revtex4-2}

\pdfoutput=1

\usepackage{graphicx}
\usepackage{dcolumn}
\usepackage{mathtools}
\usepackage{scalerel}
\usepackage{bm}
\usepackage{upgreek}
\usepackage{mathrsfs}

  \usepackage{array}
  \usepackage{verbatim}
  \usepackage{amstext}
  \usepackage{amsmath}
  \usepackage{amssymb}
  \usepackage{slashed}
  \usepackage[bbgreekl]{mathbbol}
  \usepackage{indentfirst}

 \usepackage[T1]{fontenc}
 \usepackage[latin9]{inputenc}
 \usepackage{color}
 \PassOptionsToPackage{normalem}{ulem}
 \usepackage{ulem}
 \usepackage[unicode=true,pdfusetitle,
  bookmarks=true,bookmarksnumbered=false,bookmarksopen=false,
  breaklinks=false,pdfborder={0 0 1},backref=false,colorlinks=true]
  {hyperref}
 \hypersetup{
  citecolor=blue}

\def\q2{q^2}
\def\Q2{Q^2}

\def\axion{a}
\def\gagg{g_{\axion \gamma \gamma}}
\def\gaee{g_{\axion e e}}
\def\ma{m_{\axion}}
\def\va{v_{\axion}}

\def\Ea{E_{\axion}}
\def\eVee{\rm eV_{\rm ee}}
\def\keVee{\rm keV_{\rm ee}}
\def\MeVee{\rm MeV_{\rm ee}}
\def\Ga2g{\Gamma^V_{\axion \gamma \gamma}}
\def\IPel{{\rm IP}_{el}}
\def\IPex{{\rm IP}_{ex}}
\def\IPion{{\rm IP}_{ion}}

\makeatother

\begin{document}

\title{
Inverse Primakoff Scattering for
Axionlike Particle Coupling 
}

\newcommand{\as}{Institute of Physics, Academia Sinica,
Taipei 11529, Taiwan.}
\newcommand{\bhu}{Department of Physics, Institute of Science,
Banaras Hindu University,
Varanasi 221005, India.}
\newcommand{\cusb}{Department of Physics,
School of Physical and Chemical Sciences,
Central University of South Bihar, Gaya 824236, India}
\newcommand{\ntu}{
Department of Physics, CTP and LeCosPA, National Taiwan University,
Taipei 10617, Taiwan.}
\newcommand{\ncts}{
Physics Division, National Center for Theoretical Sciences, 
National Taiwan University, Taipei 10617, Taiwan.}
\newcommand{\ndhu}{
Department of Physics, National Dong Hwa University,
Shoufeng, Hualien 97401, Taiwan.}
\newcommand{\umt}{
D\'{e}partement de Physique, Universit\'{e} de Montr\'{e}al, 
Montr\'{e}al H3C 3J7, Canada.}

\newcommand{\corrjwc}{jwc@phys.ntu.edu.tw}
\newcommand{\corrcpl}{cpliu@mail.ndhu.edu.tw}
\newcommand{\corrhw}{htwong@phys.sinica.edu.tw}

\author{ C.-P.~Wu }  \affiliation{ \umt }
\author{ C.-P.~Liu }  \altaffiliation{ \corrcpl } \affiliation{ \ndhu } \affiliation{ \ncts }
\author{ Greeshma~C. } \affiliation{ \as } \affiliation{ \cusb }
\author{ L.~Singh }  \affiliation{ \as } \affiliation{ \cusb }
\author{ J.-W.~Chen } \altaffiliation{ \corrjwc }  \affiliation{ \ntu } \affiliation{ \ncts }
\author{ H.-C.~Chi} \affiliation{ \ndhu }
\author{ M.K.~Pandey }  \affiliation{ \ndhu }, \affiliation{ \ntu }
\author{ H.T.~Wong } \altaffiliation{ \corrhw } \affiliation{ \as }

\date{\today}

\begin{abstract}

Axionlike particles (ALPs) can be produced in the Sun, and
are considered viable candidates for the cosmological dark matter (DM). 
It can decay into two photons or interact with matter. 
We identify new inelastic channels of
inverse Primakoff processes due to atomic excitation
and ionization. 
Their cross sections are derived by incorporating full electromagnetic
fields of atomic charge and current densities, and computed 
by well-benchmarked atomic many-body methods. 
Complementing data from the underground XENONnT and 
surface TEXONO experiments are analyzed. 
Event rates and sensitivity reaches are evaluated
with respect to solar- and DM-ALPs. 
New parameter space  
in ALP couplings with the photons versus ALP masses 
in (1~eV$-$10~keV) not previously accessible to laboratory experiments 
are probed and excluded with solar-ALPs.
However, at regions where DM-ALPs have already decayed,
there would be no ALP-flux and hence no interactions
at the detectors in direct search experiments.
No physics constraints can be derived.
Future projects would be able to evade the stability bound
and open new observable windows  
in (100~eV$-$1~MeV) for DM-ALPs.

\end{abstract}


\maketitle

\section{Introduction}

Axions are hypothetical particles first introduced
to solve the strong CP problem with the spontaneous breaking of the
Peccei-Quinn symmetry~\citep{Peccei:1977hh,Peccei:1977ur,Weinberg:1977ma,Wilczek:1977pj}.
Theoretical and experimental studies later evolved from the original
``QCD axions'' to variants generically called ``axionlike particles''
(ALPs, denoted as $a$), whose masses and coupling strengths with matter are no longer
related. 

Sources of ALPs are diverse: 
they are well-motivated dark matter (DM)
candidates~\cite{PRESKILL1983127,ABBOTT1983133,DINE1983137}, 
and can be produced in astrophysical environments and
terrestrial laboratories. 
Measurable signatures can be made under
a wide variety of experimental techniques~\citep{RAFFELT1982323,Cameron:1993mr,Graham:2015ouw,
Irastorza:2018dyq,Sikivie:2020zpn,Choi:2020rgn,ParticleDataGroup:2022pth,AxionLimits},
which include micro-wave cavities, solar-ALP
helioscopes, indirect searches of anomalous electromagnetic radiations
in the universe, as well as 
production by colliders or strong lasers.
Constraints of ALP properties are also derived from cooling of astrophysical
objects. 

Data from the DM direct search experiments 
can provide constraints to 
the ALP-photon coupling $\gagg$ and 
the ALP-fermion coupling $g_{aff}$. 
Through the axio-electric effect, competitive bounds
on the ALP-electron coupling $\gaee$ 
have been set in the range of ALP masses
$40 ~ {\rm eV} {<} \ma {<} \mathcal{O}(1 ~ {\rm MeV})$
(natural units with $\hbar {=} c {=} 1$
are used throughout this article unless otherwise stated)
for DM-ALPs, and in a smaller mass range
with solar-ALPs (see Ref.~\citep{AxionLimits}). 
Laboratory constraints on $\gagg$, however, are comparatively scarce. 
So far, they are primarily derived from Bragg
scattering of solar-ALPs on crystal targets, and are only applicable to
$\ma{<}\mathcal{O}(1~{\rm keV})$~\citep{Avignone:1997th,Bernabei:2001ny,COSME:2001jci,CDMS:2009fba,Armengaud:2013rta}.

The theme of this work is 
to expand and improve the sensitivities
of laboratory experiments
in probing the $m_{a}$-$\gagg$ parameter space 
using the inelastic 
inverse Primakoff (IP) scatterings between ALPs
and matter as the detection channels.

This article is structured as follows:
Theoretical formulation of ALP interactions with matter via $\gagg$-coupling, as well as the  
evaluation of the corresponding cross sections are discussed in Sect.~\ref{sect::formulation}.
The observable events on Earth taking solar-ALPs and DM-ALPs as sources
are given in Sect.~\ref{sect::eventrates}.
Physics results on $\gagg$ by the IP processes from selected 
laboratory experiments are presented in Sect.~\ref{sect::expt},
and are compared with other experimental, astrophysical and cosmological bounds.
It will be shown that 
new limits are achieved for solar-ALPs. 
However,
the current sensitivity parameter space for DM-ALPs
is mostly forbidden for direct detection 
by the cosmological stability bound,
though next generation of experiments would be able to open a new window
and study the unexplored parameter space.



\section{Formalism}
\label{sect::formulation}

The interaction Lagrangian of an ALP field ($a$) with the photon
field ($A^{\mu}$) and the Standard-Model fermion fields ($\Psi_{f}$
with $f$ specifying its flavor) is generally written  as~\cite{ParticleDataGroup:2022pth}
\begin{equation}
\mathcal{L}_{I}=-\frac{\gagg}{4}aF_{\mu\nu}\widetilde{F}^{\mu\nu}-\sum_{f}\,\frac{g_{aff}}{2m_{f}}(\partial_{\mu}a)\overline{\Psi}_{f}\gamma^{\mu}\gamma_{5}\Psi_{f}\,,\label{eq:L_agg}
\end{equation}
where $F^{\mu\nu} {=} \partial^{\mu}A^{\nu} {-} \partial^{\nu}A^{\mu}$ and
$\widetilde{F}^{\mu\nu} {\equiv} \frac{1}{2}\epsilon^{\mu\nu\rho\sigma}F_{\rho\sigma}$
are the photon field tensor and its dual, with $\epsilon^{0123} {=} 1$;
$m_{f}$ is the fermion mass; and $\gagg$ ($g_{aff}$) denotes the
strength of the ALP-photon (ALP-fermion) coupling. These interaction
terms provide the foundation for ALP detection and production. In this
work, we will focus only on the processes resulting from a finite
$\gagg$.


 \begin{figure}
 \includegraphics[width=8.2cm]{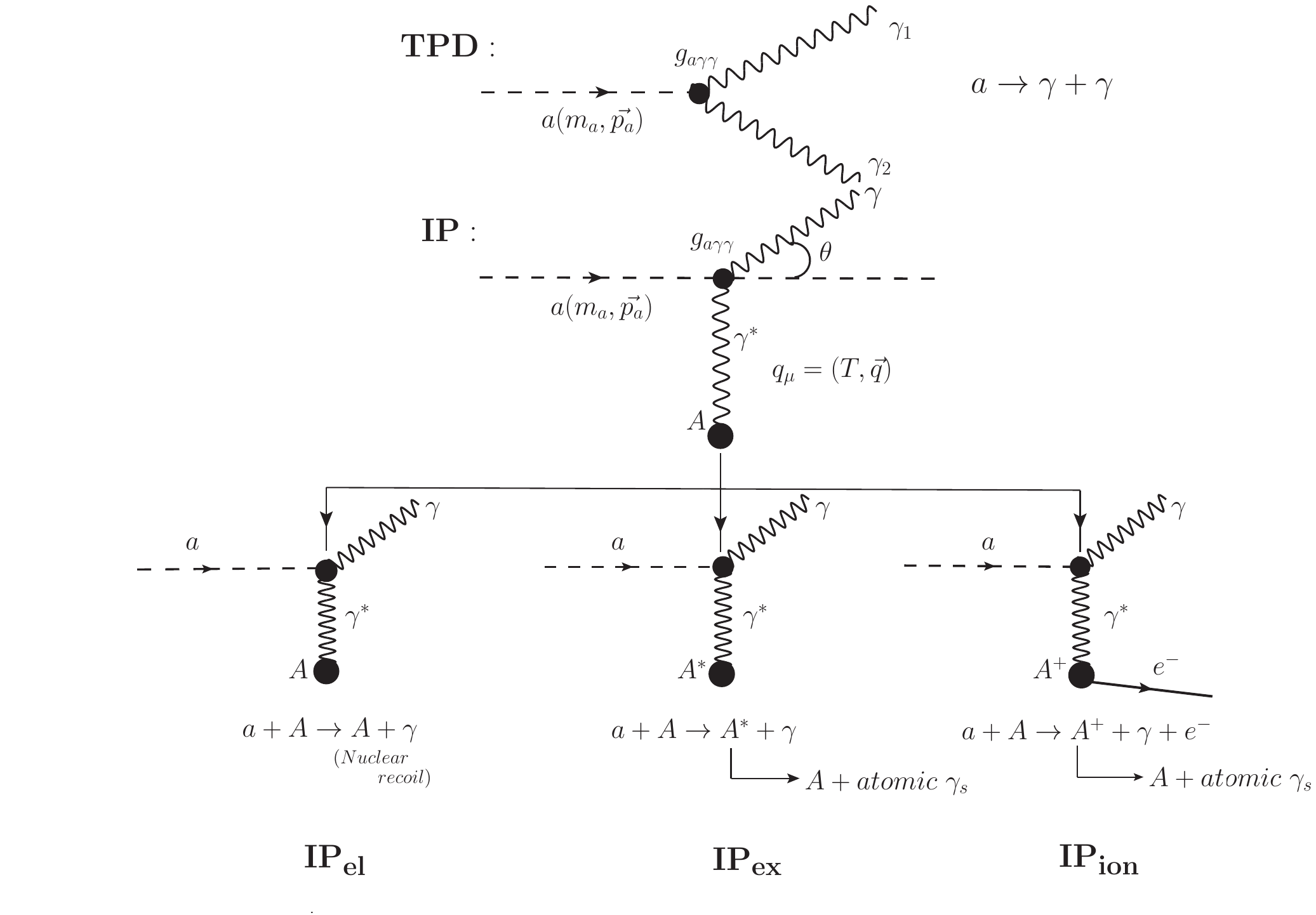}
 \caption{Schematic diagrams of ALP two-photons decay in vacuum (TPD) 
and the three IP scattering channels in matter, where kinematics allows
 one of the photons to be virtual.}
 \label{fig::gagg-schematics}
 \end{figure}


The coupling $\gagg$ can manifest experimentally in many ways as
shown in Fig.~\ref{fig::gagg-schematics}. The most straightforward channel is
where an ALP can have two-photon decay (TPD) in vacuum: 
\begin{equation}
\axion\rightarrow\gamma_{1}+\gamma_{2}~~:~~\Ga2g=\frac{1}{64\pi}\gagg^{2}\ma^{3}~~,
\label{eq::Ga2g}
\end{equation}
where $\Ga2g$ is the decay rate at rest per
ALP. When one of the photons becomes virtual in a medium and is absorbed
by the target atom $A$, it gives rise to 
IP scattering~\cite{Barshay:1981ky,Avignone:1988bv,Buchmuller:1989rb,Paschos:1993yf,Creswick:1997pg,Dent:2020jhf,Gao:2020wer,Abe:2021ocf}
with four-momentum transfer $q_{\mu}{\equiv}(T,\,\vec{q})$. There
are three IP reaction channels: $\axion{+}A{\rightarrow}$ 
\begin{equation}
\begin{cases}
~~\gamma+A & \IPel:\text{elastic scattering}\\
~~\gamma+A^{*} & \IPex:\text{atomic excitation }\\
~~\gamma+A^{+}+e^{-}~~ & \IPion:\text{atomic ionization }
\end{cases}~~.\label{eq:channel}
\end{equation}
All four channels involve full conversion of the ALPs
so that the experimental measurable is the total energy $\Ea$.
There is no interference among them, since all have experimentally
distinguishable final states $-$ two $\gamma$'s for TPD, a single
$\gamma$ for $\IPel$, a $\gamma$ plus atomic de-excitation photons
for $\IPex$, and a $\gamma$ plus an ionized electron with atomic
transition photons for $\IPion$. At $\ma$ much lower than the nucleus
mass scales (GeV), the energy depositions at detectors are electromagnetic
without complications of nuclear recoil.

\subsection{Differential Cross Sections and Rates}

Evaluation of the double differential cross sections of the ALP IP
processes builds on our earlier work of incorporating the atomic many-body
physics effects to low energy neutrino~\cite{Chen:2013lba,Chen:2014dsa,Chen:2014ypv}
and DM~\cite{Pandey:2018esq,Liu:2021avx} interactions with matter:
\begin{equation}
	\dfrac{d\sigma_\textrm{IP}}{dTd\Omega} =
\frac{\alpha \gagg^2}{16\pi} \left( \frac{\Ea-T}{v_{a} \Ea} \right)
\left[ \frac{V_{L}}{(q^{2})^{2}}\mathscr{R}_{L}+\frac{V_{T}}{(Q^{2})^{2}}\mathscr{R}_{T} \right] \, ,
\label{eq:ddcs}
\end{equation}
where $\alpha$ is the fine structure constant, 
$v_a$ is the ALP velocity,
$q^{2}$ and $Q^{2} {=} T^{2} {-} q^{2}$ are 
the three- and four-momentum transfer squared,
respectively. 
The kinematic factors $V_L$ and $V_T$ will be given in 
Eqs.~(\ref{eq:V_L})\&(\ref{eq:V_T}), respectively.
The full unpolarized atomic response consists of the
longitudinal and transverse components, $\mathscr{R}_{L}$ and $\mathscr{R}_{T}$,
both being functions of $T$ and $q^{2}$ given by:
\begin{eqnarray}
	\mathscr{R}_{L,T} & = & \sum_{F}\overline{\sum_{I}}|\left<F|\rho,\vec{j}_{\perp}|I\right>|^{2}\,\delta\left(E_{I}-E_{F}-T\right)\,,\label{eq:resp_func}
\end{eqnarray}
which arise from the charge $\rho$ and transverse current density $\vec{j}_{\perp}$,
respectively. Note that the longitudinal current density is effectively included
in the former by current conservation and also manifested in the change
of the photon propagator from $1/Q^{2}$ to $1/q^{2}$. The initial state
$| I \rangle$ 
corresponds to the ground state of the target atom, while
the choice of final state $| F \rangle$ 
depends on the IP interaction
channels: the ground, excited and continuum states for $\IPel$,
$\IPex$ and $\IPion$, respectively.
In general, the transverse response is less than  
the longitudinal response by a factor of ${\sim} \alpha$
since the atomic current density is suppressed 
relative to the atomic charge density
by the electron velocity. 

Experimentally, the observables are
the energy of the IP photon ($E_{\gamma}$)
and the energy transferred ($T$) to the scattered atom as
	atomic recoil, excitation, or ionization. 
The differential rate recorded by a detector is with respect to
the total observable energy $\Ea$ of the ALP,
which is the sum of $E_{\gamma}$ and $T$. 
The double differential rate is with respect
to $E_{\gamma}$ and $T$ simultaneously, where more
elaborated analyses are necessary to correlate these two signals. 
We focus in this work the simplest one: 
a single differential rate with respect to $E_{a}$
	\begin{equation}
		\frac{dR_\textrm{IP}}{dE_{a}}=N_{A}\sigma_{\textrm{IP}}(E_{a})\frac{d\phi}{dE_{a}}\,,\label{eq:dR/dEa}
	\end{equation}
	where $N_{A}$ is the total number of target atoms, $\sigma_{\textrm{IP}}(E_{a})$
	and $d\phi/dE_{a}$ are the total IP cross section (a double integration
	of Eq.~(\ref{eq:ddcs})) and energy spectrum of ALPs with incident energy
	$E_{a}$, respectively.

\subsection{Regularization of Divergences}

To fully exhibit the pole structure of the photon propagator, which
is crucial in cross section calculations, the corresponding kinematic
factors $V_{L}$ and $V_{T}$ are cast in powers of $q^{2}$ and $Q^{2}$:
\begin{eqnarray}
V_{L} & = & 2 \Big[ \Ea^{2} - \ma^{2} +( \Ea - T )^{2} \Big] \q2 -( \q2 )^{2} \nonumber \\
 &  & -( T^{2}-2 \Ea T+ \ma^{2} )^{2}\,,\label{eq:V_L} \\
V_{T} & = & \ma^{4} + \frac{Q^{2}}{2 \q2} \Big[ ( \ma^{4}-4 \ma^{2} \Ea T ) \nonumber \\
 &  & +( 2 \ma^{2} + 4 \Ea^{2} - 4\Ea T+ 2T^{2} ) \Q2 - ( \Q2 )^{2} \Big] \,.
\label{eq:V_T}
\end{eqnarray}
The familiar Coulomb pole $q^{2} {=} 0$ is realized in the longitudinal
component only at the forward angle $\theta {=} 0$ along with the condition
$T {=} E_{a}(1 {-} v_{a})$. 
Since the last term of Eq.~(\ref{eq:V_L})
vanishes identically as $q^{2} {\rightarrow} 0$, only a single pole 
in $V_{L}/(q^{2})^{2}$ is produced. 
This divergence is usually regulated
by Coulomb screening in 
media (see, for example, Refs.~\cite{Avignone:1988bv,Buchmuller:1989rb,Marsh:2014gca,Dent:2020jhf,Gao:2020wer})
via changing the longitudinal propagator from $1/q^{2}$ to $1/(q^{2}+\Lambda_{L}^{2})$.
The cutoff $\Lambda_{L}$, which is medium-dependent, modifies the
infinite-range Coulomb interaction to a Yukawa one with a range of
order $1/\Lambda_{L}$, typically ${\sim} \text{\r{A}}$. 
We note that this procedure only applies to electrons which are not localized,
for example, valence electrons in semiconductors. For a single atom,
its neutrality automatically screens its Coulomb field. A manifestation
of this fact is the response $\mathscr{R}_{L}$ starts at the order
of $( q^{2} )^2$, as demonstrated in Eq.~(10) and Table 1 of Ref.~\cite{Abe:2021ocf}.
(The $\mathscr{R}_L$ responses for $\IPex$ and $\IPion$ 
also start at the order of
$( q^2 )^2$, but they are due to wave function orthogonality.)
Furthermore, the inner-shell electrons
experience and contribute to screening at shorter length scales than the size
of an atom. As a result, applying a universal screening length ${\sim} \text{\r{A}}$
tends to overestimate the atomic Coulomb field. This is the main reason
that previous calculations using ${\sim} \text{\r{A}}$ screening lengths
over-predicted $\IPel$ cross sections by solar-ALPs, as was first
pointed out and corrected through realistic atomic calculations in
Ref.~\cite{Abe:2021ocf}.

\begin{figure}
\begin{center}
\includegraphics[width=8.2cm]{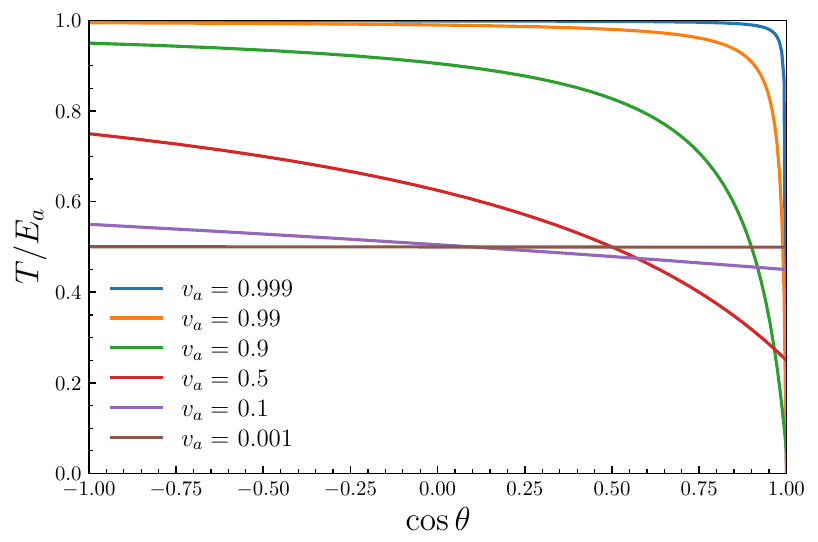} 
\end{center}
\caption{
Contours of $\Q2 {=} 0$ in the IP processes traced by the ALP scattering
angle $\cos\theta$ and fraction of energy transfer $(T/\Ea)$ for
selected values of $\va$. 
}
\label{fig::gagg-pole}
\end{figure}

The transverse component $V_{T}/(Q^{2})^{2}$ does exhibit a double
pole structure at $Q^{2} {=} 0$ for nonzero 
$m_{a}$.\footnote{Note that the $1/q^{2}$ factor in $V_{T}$ does not yield a pole,
as the numerator that follows also vanishes identically with $Q^{2} {=} T^{2}$
when $q^{2} {\rightarrow} 0$.} 
The kinematics of the incoming ALP and outgoing photon makes it possible
to have
\begin{equation}
Q^{2}=m_{a}^{2}-2E_{a}(E_{a}-T)(1-v_{a}\cos\theta)\,,\label{eq:Q^2}
\end{equation}
vary from time-like to space-like as the scattering angle ($\theta$) increases.
Fig.~\ref{fig::gagg-pole} depicts the $Q^{2} {=} 0$ contours traced by
$\cos\theta$ and the fraction of energy transfer
$(T/\Ea)$ at selected $v_{a}.$ For ultra-relativistic (UR) ALPs with
$v_{a} {\approx} 1$, such as solar-ALPs, the divergence only happens
at the forward angle. As $v_{a}$ becomes less relativistic, 
the time-like to space-like transition happens at some intermediate scattering angle.
For non-relativistic (NR) DM-ALPs with $v_{a} {\sim} 10^{-3}$ (where $E_{a} {\approx} m_{a}$),
the pole is realized at a tiny range around $T {\approx} m_{a}/2$. 
As the virtual photon is absorbed by the target, 
the kinematics of the final state of the target further limits 
the available $\Q2$-space. 
In general, $\Q2$ can be both space-like and time-like for
$\IPex$ and $\IPion$, while
$\Q2 {\le} 0$ for $\IPel$.



To regulate the transverse photon pole, at which the virtual photon
approaches the real limit, we follow the approach of Ref.~\cite{Chen:2016lyr}
and modify the photon propagator according to the complex refractive
index of the detector material: $\tilde{n} {=} n_{r} {+} in_{i}$, where the
real part $n_{r}$ gives the normal refractive index, and the imaginary
part is related to the attenuation coefficient $\Lambda_{T} {=} 2Tn_{i}$.
If the detector consists of only one type of atom with a number density
$n_{A}$ and photo-absorption cross section $\sigma_{\gamma}(T)$,
we can further relate $\Lambda_{T} {=} n_{A}\sigma_{\gamma}(T)$. 
As a result,
the photon propagator in  vacuum should be modified to be
\begin{align}
\frac{1}{Q^{2}}= & \frac{1}{T^{2}-q^{2}}\rightarrow\frac{1}{T^{2}\tilde{n}^{2}-q^{2}}\nonumber \\
\rightarrow & \frac{1}{(Q^{2}-\frac{1}{4}\Lambda_{T}^{2})+T^{2}(n_{r}^{2}-1)+iT\Lambda_{T}} ~~ .
\label{eq:Q2_reg}
\end{align}
For simplicity, we take $n_{r} {=} 1$ in this paper, which is 
a good approximation for X-ray and $\gamma$-ray photons, and leave refined treatments
for photons of lower or near-resonance energies for future studies.
The square of the propagator in Eq.~(\ref{eq:ddcs}), 
which itself is an absolute value squared, becomes
\begin{equation}
\frac{1}{(Q^{2})^{2}}\rightarrow\frac{1}{(Q^{2}-\Lambda_{T}^{2}/4)^{2}+T^{2}\Lambda_{T}^{2}}\,.\label{eq:Q4_reg}
\end{equation}
Typical values are $n_{A} {\sim} 10^{22}/\textrm{cm}^{2}$, 
and $\sigma_{\gamma}(T) {\lesssim} 10^{6}\,\mathrm{barn}$
for $T {>} 100\,\mathrm{eV}$. As a result, $\Lambda_{T}$ is not larger
than a few eV, so the shift in the pole position is not significant,
and $T\Lambda_{T}$ is within the range $10^{2} {-} 10^{4}\,\mathrm{eV}^{2}$.
We note that the above treatment requires the virtual photon to be
absorbed inside the detector, or equivalently the length dimension of the
detector has to be substantially bigger than the attenuation length
$1/\Lambda_{T}$.


\subsection{Equivalent Photon Approximation \label{subsec:EPA}}

To acquire further insight into the transverse contribution, it is useful
to perform an equivalent photon approximation (EPA), similar to the
method adopted in Ref.~\cite{Chen:2016lyr}, by 
(i) setting $V_{L} {=} 0$, 
(ii) keeping the sole $m_{a}^{4}$-term in $V_{T}$, and 
(iii) substituting the transverse response $\mathscr{R}_{T}(T,q)$ by photo-absorption
cross section -- $\mathscr{R}_{T}(T,q)\approx T\sigma_{\gamma}(T)/(2\pi^{2}\alpha)$.
The single differential cross section (SDCS) for $\IPion$ can then
be easily integrated to give 
\begin{eqnarray}
\left.
\dfrac{d\sigma}{dT}\right|_{\IPion}^{\textrm{EPA}}
 & ~ = ~  & 
\dfrac{g_{a\gamma\gamma}^{2}}{32\pi^{2}}
\left( \frac{\sigma_{\gamma}}{\Lambda_{T}} \right) 
\left( \frac{m_{a}^{4}}{v_{a}^{2}E_{a}^{2}} \right) \times \\
& & 
~~~~~~~~ 
\tan^{\mbox{-}1}\left[\dfrac{Q^{2}-\Lambda_{T}^{2}/4}{T\Lambda_{T}}\right]\bigg|_{Q_{\min}^{2}}^{Q_{\max}^{2}}
\, .   \nonumber
\label{eq:dcs_EPA}
\end{eqnarray}

Several important features can be observed:
First, the $1/\Lambda_{T}$ dependence clearly traces the divergence
resulting from the double pole. Together with $\sigma_{\gamma}$,
it leads to a single-atom cross section suppressed by $1/n_{A}$,
due to the fact that the effective interaction range becomes shorter
as the medium gets denser. 
Second, the factor 
$\ma^{4}/(v_{a}^{2} \Ea^{2}){=} \ma^{2} (1 {-} v_{a}^{2})/v_{a}^{2}$
indicates that this contribution favors NR ALPs with a big mass. 

Moreover, when both $Q_{\max}^{2}$ and $-Q_{\min}^{2}$ are much
larger than $T\Lambda_{T}$, the arc-tangent value saturates to $\pi$.
As a result, the SDCS becomes a $T$-independent constant as long
as the kinematics allows $Q^{2}$ to change sign when the scattering angle
increases. For highly NR cases with $v_{a} {\ll} 1$, this requires
$T$ in the range of $m_{a}(1 {\pm} v_{a})/2$, 
and, as an approximation, the total cross section 
can be easily integrated to be:
\begin{equation}
\sigma_{\IPion,\textrm{NR}}^{\textrm{EPA}}
~ \approx ~ 
\dfrac{g_{a\gamma\gamma}^{2}}{32\pi}
\left( \frac{1}{n_{A}} \right)
\left( \frac{m_{a}^{3}}{v_{a}} \right) \,.
\label{eq:sigma_EPA}
\end{equation}

Considering a detector of volume $V$ (the number of atoms is thus $n_{A}V$)
and an ALP number density of $n_{a}$ (the flux is thus $n_{a}v_{a}$),
the event rate will be 
\begin{equation}
R_{\IPion,\textrm{NR}}^{\textrm{EPA}} 
~ \approx ~  
\dfrac{g_{a\gamma\gamma}^{2}}{32\pi}m_{a}^{3}(n_{a}V)
~ = ~
2\Ga2g\,(n_{a}V) \, .
\label{eq:R_IPion_EPA}
\end{equation}
Note that $n_{a}V$ is the total number of ALP inside the detector
volume, so this result gives the total $\IPion$ event rate twice
as much as the two-photon decay (time dilation is negligible because
$v_{a} {\ll} 1$). The factor of two can be understood in the context
of EPA that either of the decayed photons can play the role of the
intermediate state in the IP process. 
This result might seem surprising as the predicted rate
only depends on a detector volume, but not on the target characteristics.
However, one should be alert that this approximation is only valid
when all the underlying assumptions hold true. 
For instance, if ALP is not extremely NR, 
or when the transverse contribution is
sub-dominant, the cross sections and rates have to be explicitly computed.
We will discuss this in more detail through concrete cases 
in the next section.



\begin{figure}
	\includegraphics[width=8.2cm]{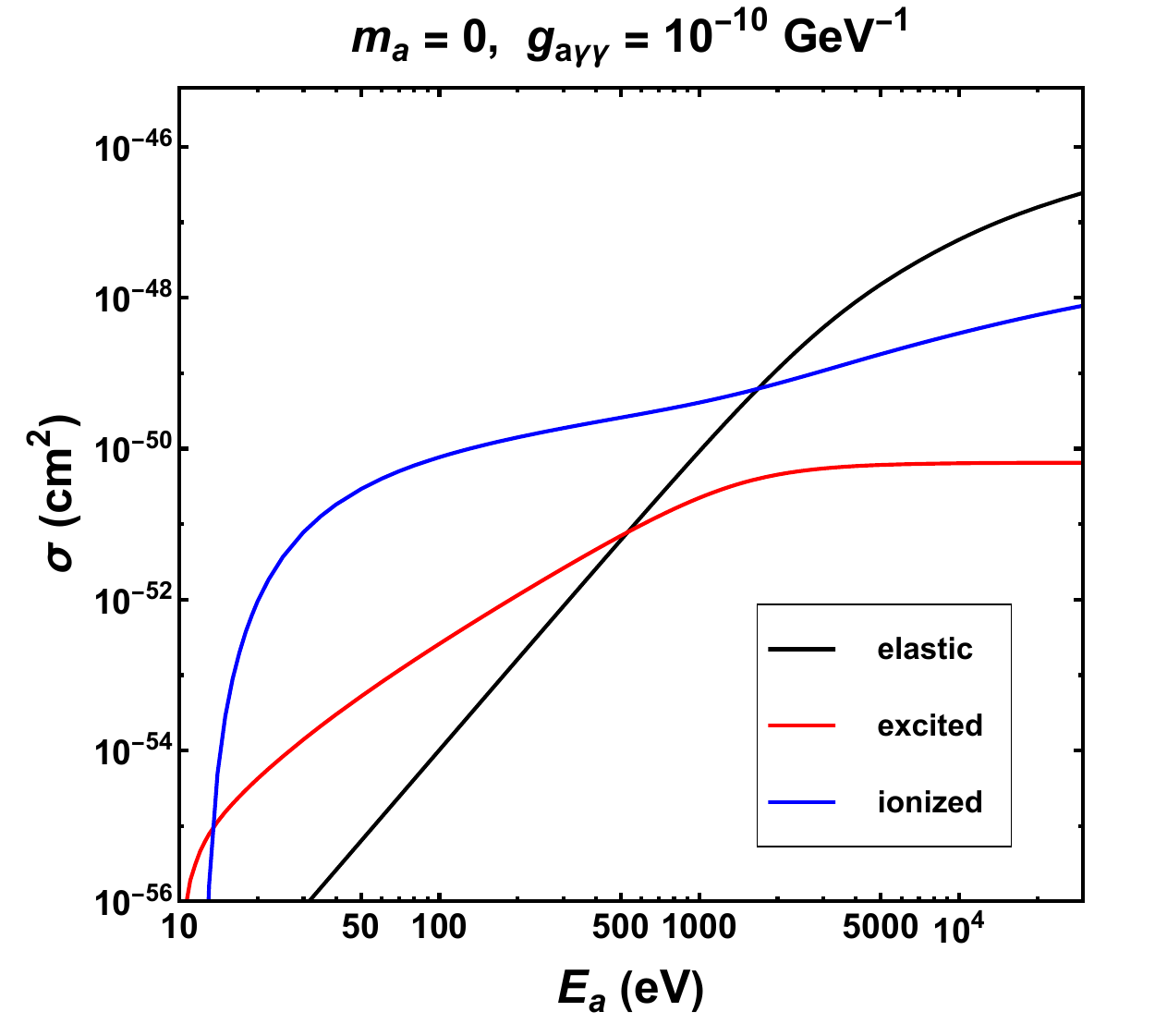}\\
	\caption{ Total cross sections for the three IP detection
		channels for the case of a massless ALP scattering off a xenon atom with
		$\gagg{=}10^{\mbox{-}10}~{\rm GeV^{\mbox{-}1}}$.
	}
	\label{fig:TCS-IP}
\end{figure}


\subsection{Selected Numerical Results \label{sec:IID}}

In this subsection, we present numerical results using xenon as the target and assuming 
$\gagg {=} 10^{-10}\textrm{GeV}^{\mbox{-}1}$, 
to illustrate several key kinematic features 
when ALPs are either UR or NR.
  
The scattering cross sections for a massless ALP as a function of $\Ea$ 
are displayed in Fig.~\ref{fig:TCS-IP}, where  
the black, red, blue curves denote the $\IPel , \IPex {\rm ~ and ~} \IPion$
channels, respectively. 
At high energy where $\Ea {\gtrsim} 2\,\textrm{keV}$, 
$\IPel$ dominates the total cross section. 
On the other hand, $\IPion$ provides the leading contribution at $\Ea {<} 1 ~ {\rm keV}$, 
where $\IPel$ is suppressed due to the large electronic screening in the charge form factor. 

\begin{figure}
\textbf{(a)}\\
 \includegraphics[width=8.2cm]{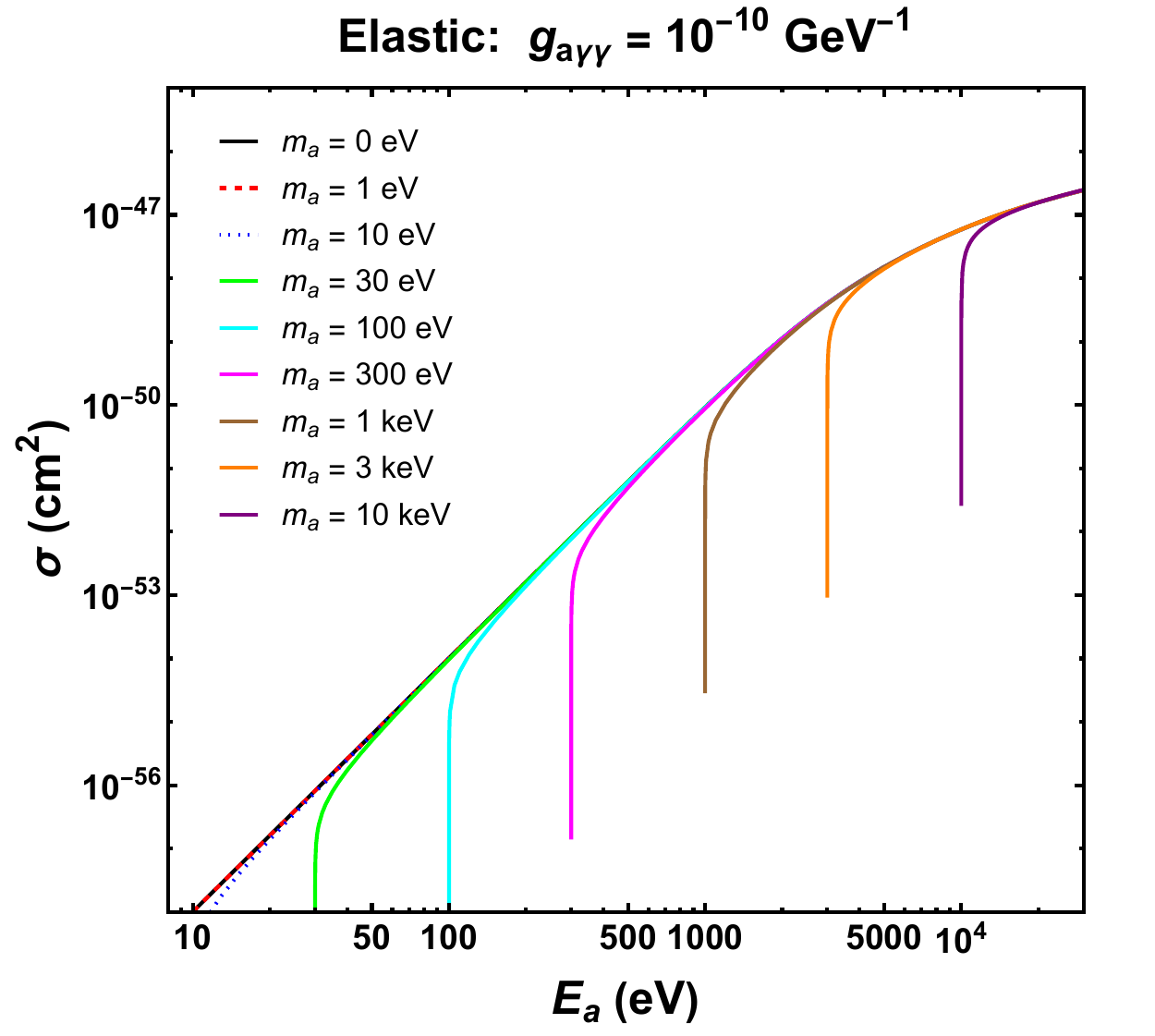}\\
\textbf{(b)}\\
\includegraphics[width=8.2cm]{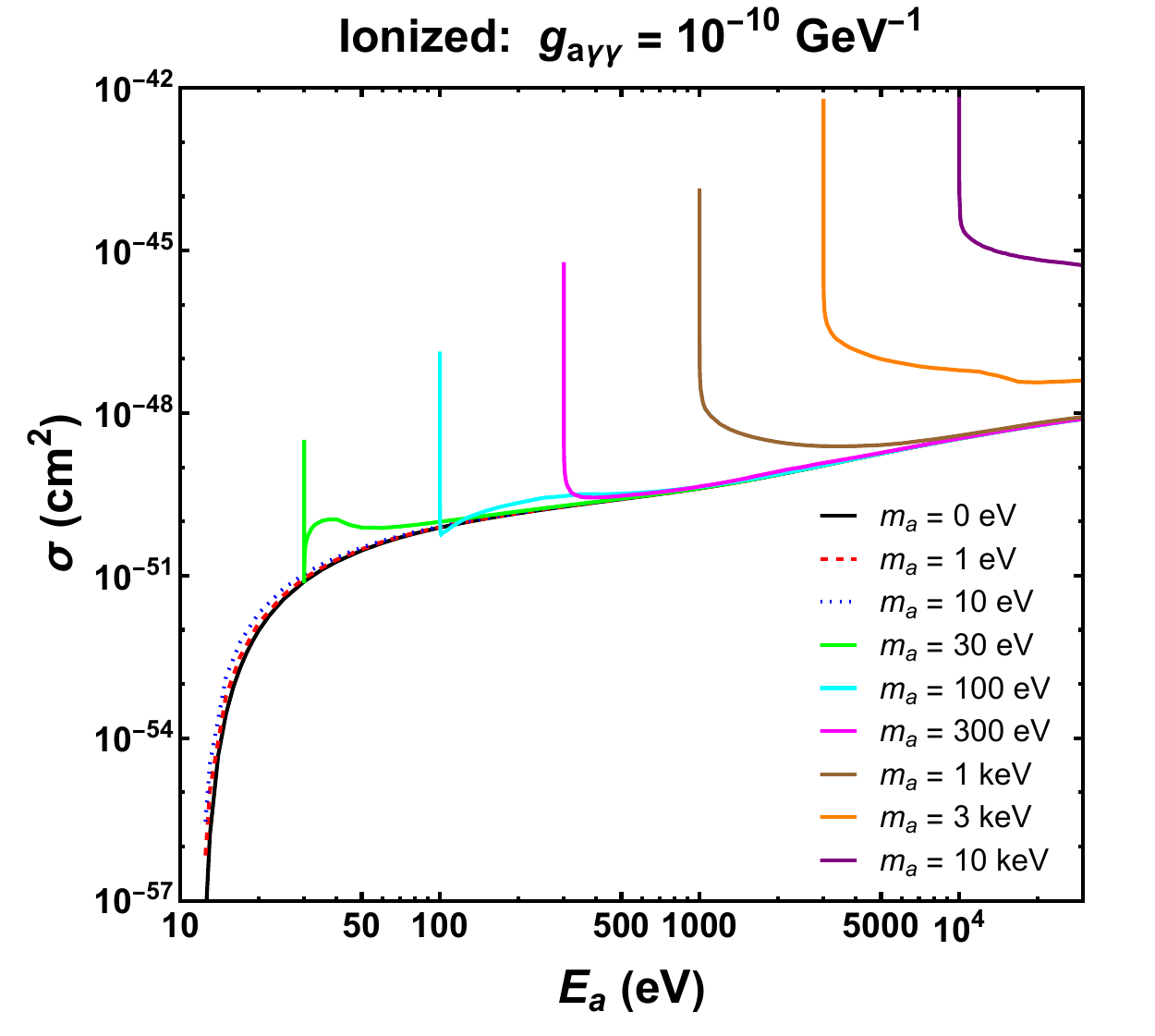} 
\caption{ Total cross sections for massive ALPs scattering off a xenon atom in the (a) elastic $\IPel$ and 
(b) ionization $\IPion$ channels, at $\gagg{=}10^{\mbox{-}10}~{\rm GeV^{\mbox{-}1}}$.
}
\label{fig:TCS-mass} 
\end{figure}

Several special cases of massive ALPs 
are shown in Figs.~\ref{fig:TCS-mass}(a)\&(b) 
for $\IPel$ and $\IPion$, respectively. 
In $\IPel$, $Q^2$ can only be space-like. 
As illustrated in Fig.~\ref{fig:TCS-mass}(a),
the total cross section is mostly due to the longitudinal response, 
and approaches the massless limit whenever the ALPs become relativistic. 
Conversely, the cross section drops quickly to zero near energy thresholds due to the NR suppression of  
$V_L {\propto} q^2$ and the vanishing of phase space. 
On the other hand, Fig.~\ref{fig:TCS-mass}(b) indicates 
that the contributions from $\IPion$ are complicated by the enhancement of the double pole in $\Q2$,
which is associated with the transverse response and is proportional to $m_{a}^{2}(1-v_{a}^{2})^{2}/v_{a}^{2}$. 
The departures from the massless case are more dramatic near the energy thresholds, and also grow with the mass.   

\begin{figure}
\textbf{(a)}\\
 \includegraphics[width=8.2cm]{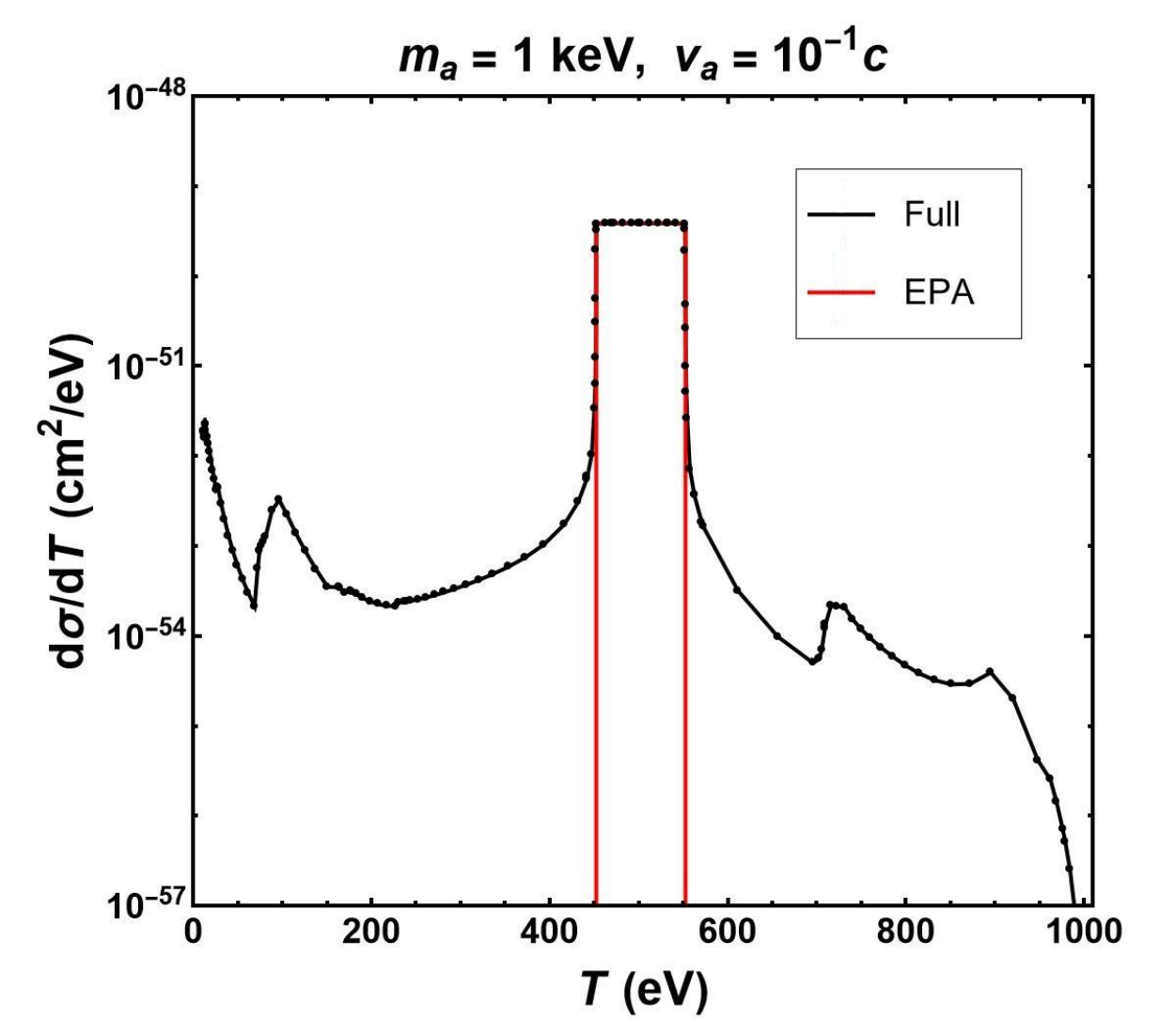}\\
\textbf{(b)}\\
 \includegraphics[width=8.2cm]{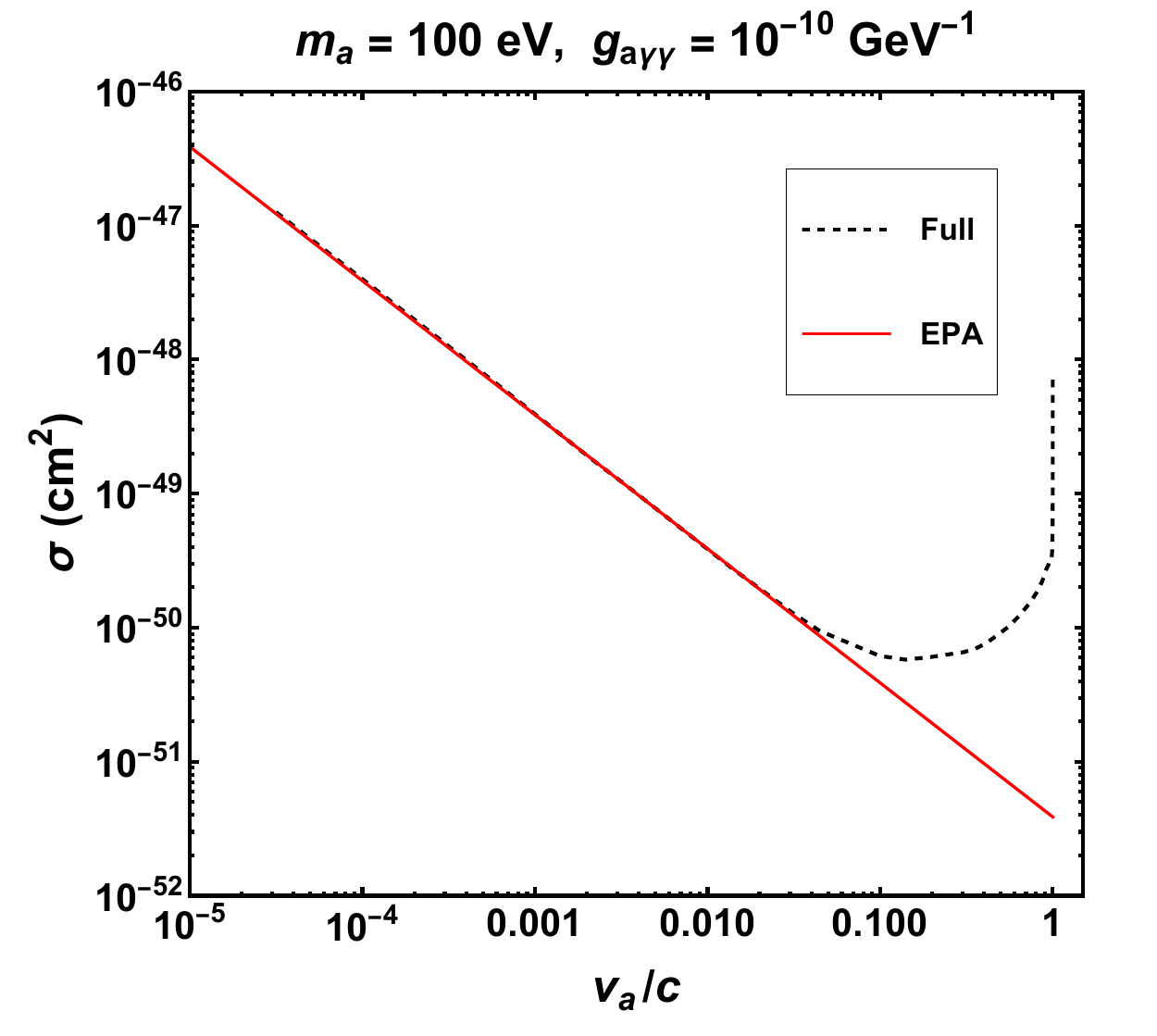}
\caption{ 
Comparison between the full calculation and EPA 
for the $\IPion$ process in xenon:
(a) the differential cross section with $\ma{=}1~{\rm keV}$ and 
(b) total cross section with $\ma{=}100~{\rm eV}$, 
with $\gagg{=}10^{\mbox{-}10}~{\rm GeV^{\mbox{-}1}}$ in both cases.
}
\label{fig:vsEPA} 
\end{figure}

The $\IPion$ SDCS for an NR ALP with $\va {=} 0.1$ and $\ma {=} 1\,\textrm{keV}$ is presented
in Fig.~\ref{fig:vsEPA}(a). 
The full calculation (in black), which is based on the atomic wave functions obtained by 
the frozen core approximation (FCA) and numerical integration of Eq.~(\ref{eq:ddcs}), 
is compared with the EPA (in red) as prescribed by Eq.~(\ref{eq:dcs_EPA}). 
In the energy range where the double pole can be accessed ($450 - 550\,\textrm{keV}$), 
the EPA is indeed an excellent approximation. 
Outside this region, where there are contributions from the longitudinal and other transverse responses, 
the corrections to the total cross section are at the ${\sim} 1\%$ level. 
Fig.~\ref{fig:vsEPA}(b) shows the total $\IPion$ cross section 
for ALP at $\ma {=} 100 ~ {\rm eV}$ 
as a function of $\va$. 
The EPA works well in the NR regime of $\va {\lesssim} 0.05$. 
As $\va$ approaches the relativistic limit, 
contributions from the longitudinal responses gradually take over and become dominant. 


\section{Source-Specific Event Rates}
\label{sect::eventrates}

Two ALP sources are considered in this work: 
one from the Sun, and the other under the assumption that all the galactic dark matter are ALPs.


\subsection{Solar-ALPs}


\begin{figure}
 \includegraphics[width=8.2cm]{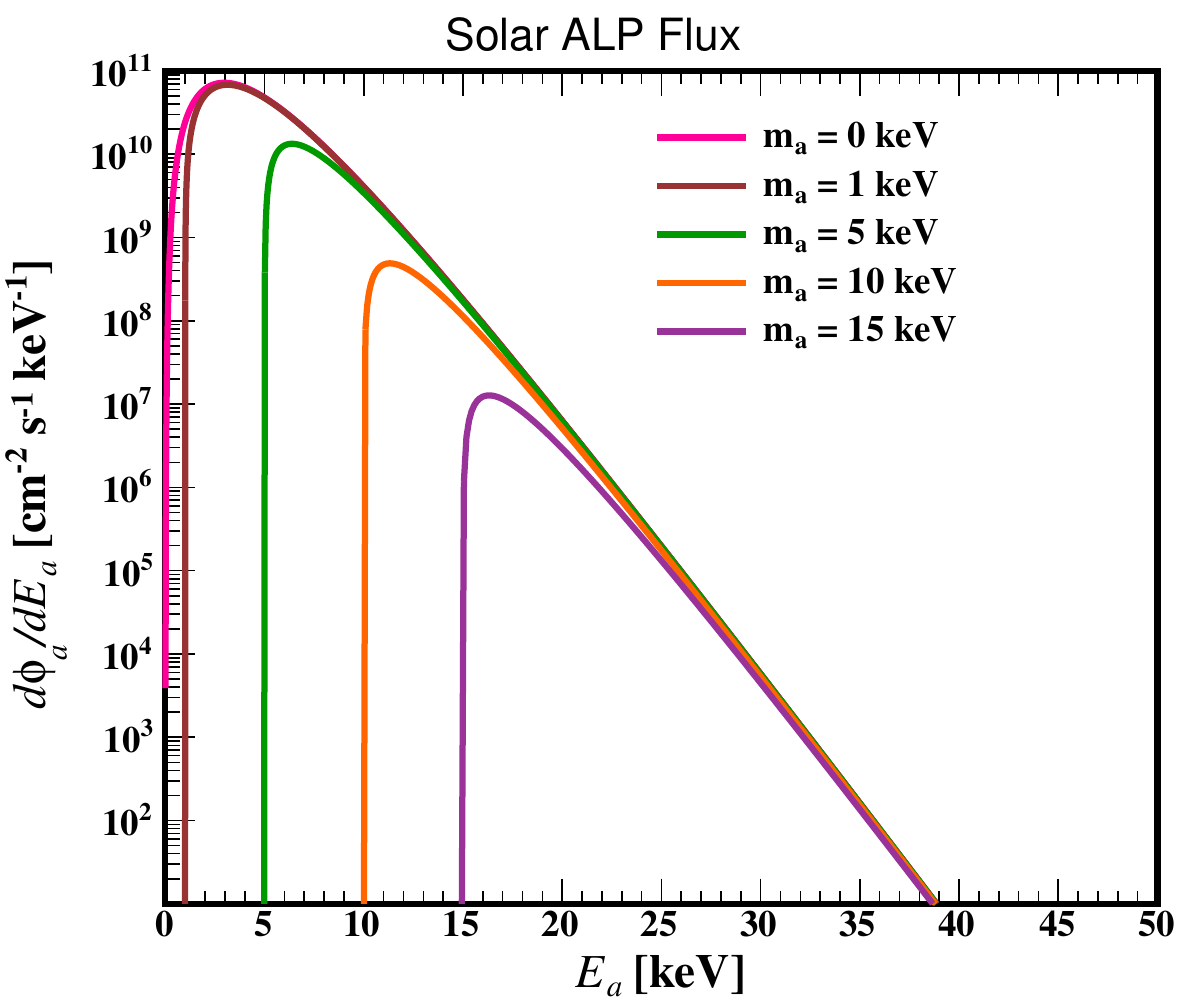}\\
\caption{  
The differential spectrum for solar-ALPs,  
following Ref.~\cite{Lella::2000}
and taking $\gagg{=}10^{\mbox{-}10}~{\rm GeV^{\mbox{-}1}}$
and several $\ma$ as illustration.
}
\label{fig:solar-ALP-flux} 
\end{figure}


The Sun can produce ALPs through both $\gagg$ and $\gaee$ processes. 
Since only detection channels involving $\gagg$ are considered in this work,
we take only the solar-ALP flux generated through the Primakoff process~\cite{solar-ALP-spectra}. 
The resulting detectable event rate is consequently proportional to $\gagg^4$.  

With the Primakoff process,
thermal photons in the solar interior are converted
into ALPs in the Coulomb fields of nuclei and electrons
in the solar plasma. The recoil effect is ignored, and
the axion energy $\Ea$ is identical to the photon energy. The
conversion rate is given by~\cite{Lella::2000}: 
\begin{eqnarray}
	\Gamma_{\gamma \rightarrow a} &= &\frac{g_{a \gamma \gamma}^{2} \textrm{T}_p \kappa^{2}}{32 \pi}\nonumber\\
	&& \times  \Bigg[\frac{ (m_{a}^{2} - \kappa^{2})^{2}  + 4 E_{a}^{2}\kappa^{2}}{4 E_{a} p_a \kappa^{2}} \ln\left( \frac{(E_{a} + p_a)^{2} + \kappa^{2}}{(E_{a} - p_a)^{2} + \kappa^{2}} \right) \nonumber\\
	&& - \frac{m_{a}^{4}}{4 E_{a} p_a \kappa^{2}} \ln\left(\frac{(E_{a} + p_a)^{2}}{(E_{a} - p_a)^{2}}  \right)  -1 \Bigg]\,,
\end{eqnarray}
where $p_a $ is the ALP momentum. 
In the UR case where $p_a {\approx} E_a$, the formula can be simplified to a more popular form~\cite{Reffelt::1988}:
\begin{equation}
\Gamma_{\gamma \rightarrow a}^{\textrm{UR}} = \frac{g_{a \gamma \gamma}^{2} \textrm{T}_p \kappa^{2}}{32 \pi} \left[ \left( 1 + \frac{\kappa^{2}}{4 \Ea^{2}}\right) \ln \left( 1 + \frac{4 \Ea^2}{\kappa^{2}} \right) -1 \right]\,. 
\end{equation}
where $\textrm{T}_p$ is the plasma temperature, and $\kappa$ is the Debye screening scale given by~\cite{Reffelt::1986}  
\begin{equation}
  \kappa^{2} = \frac{4 \pi \alpha}{T}\Big( n_{e}^{eff} + \sum_{j} Z^{2}_{j}n_{j}^{eff}  \Big) ~ ,
\end{equation}
with $n_{e}^{eff}$ and $n_{j}^{eff}$ are, respectively, 
the effective number density of electrons and ions with nuclear charge $Z_{j}$. 
The approximate effective electron number density can be expressed as  
\begin{equation}
n_{e}^{eff} \simeq \frac{(X+1)}{2} \left( \frac{\rho}{m_{u}} \right) ~ ,
\end{equation}
where $X$ is the hydrogen mass fraction, $\rho$ is the mass density and $m_{u}$ is the atomic mass unit (approximately the proton mass).

The differential flux of solar-ALPs on Earth can be written as
\begin{equation}
\frac{d\phi_{a}}{d \Ea} = \frac{1}{\pi^{2}d^{2}_{\odot}}
\int_{0}^{R_{\odot}} dr ~
\Gamma_{\gamma \rightarrow a}  ~
\left[ ~ \frac{ r^{2} ~ \Ea^{2}}{e^{\Ea / T_p} -1} ~ \right]
\end{equation}
where $R_{\odot}$ is the solar radius,
$d_{_\odot} {=} 1.5 {\times} 10^{13} ~{\rm cm}$ 
is the distance to the Sun, and the integration is with respect to
the radial distance from the solar center ($r$). 
In order to perform the integration, we adopted the BS05-AGSOP solar model \cite{Bahcall::2005} 
for the radial profiles of ${T}_p$ and $\kappa$. 
The solar-ALP spectra thus derived are depicted in 
Fig.~\ref{fig:solar-ALP-flux}.

Once the solar-ALP spectra (${d\phi_a}/{dE_a}$) are derived,
the event rates of the three IP channels can be evaluated 
directly by Eq.~(\ref{eq:dR/dEa}). 
However, in the derivation of the TPD rate, 
it is important to take the time dilation factor
into account since most solar-ALPs are relativistic, so that the TPD rate
\begin{equation}
	\dfrac{dR_{\textrm{TPD}}}{d\Ea} = (\gamma^{-1}\Ga2g )~ \left( \dfrac{V}{\va} \right) ~\dfrac{d\phi_{\axion}}{d\Ea} 
	\label{eq::dRdE_TPD}
\end{equation}
carries an additional correction factor of $\gamma^{-1} {=}  m_a/E_a$ to the decay rate at rest.

\subsubsection{Solar-ALPs at sub-keV mass (ultra-relativistic)}

As displayed in Fig.~\ref{fig:solar-ALP-flux}, 
the typical energy of solar-ALP is in the range of a few to tens of keV. 
Therefore, ALPs produced in the Sun with sub-keV mass are mostly relativistic 
and, as explained in Sect.~\ref{sec:IID}, the atomic longitudinal response dominates the IP processes. 
The differential rates expected at a liquid xenon detector with a ton-year exposure
for the four observation channels at $\gagg {=} 10^{\mbox{-}10}\,\textrm{GeV}^{\mbox{-}1}$
and $\ma {=} 0.1~{\rm keV}$
are displayed in Fig.~\ref{fig:dR/dT_solar}(a). 
The gross feature of the three IP channels is consistent with 
Figs.~\ref{fig:TCS-IP}\&\ref{fig:TCS-mass} --
in the entire keV range, the $\IPel$ has the largest count rates except when $\Ea {\lesssim} 2\,\textrm{keV}$, 
where $\IPion$ starts to play an important role.
On the contrary, $\IPex$ does not contribute significantly, 
and TPD is also suppressed by the time dilation factor.

The total event rates of the four channels at different $m_a$
are presented in Table~\ref{tab::Event-Rates}. 
At sub-keV energy (that is, 1~meV, 1~eV, and 1~keV), the $\IPel$-rates dominate 
and show little mass dependence as $\va {\rightarrow} 1$ in the $\Ea {\sim} {\rm keV}$ energy range. 
The sub-leading contribution is from $\IPion$ -- still substantial at 10\%-20\% level. 
In addition, the $\ma^4 / \Ea^2 {\propto} \ma^2$ factor in the
transverse response leads to the NR enhancement
in Eq.~(\ref{eq:dcs_EPA}) and provides bigger correction at higher $\ma$.
Unlike $\IPex$ which is relatively small and almost mass-independent, 
the TPD sensitively depends on $\ma^4$, following Eq.~(\ref{eq::dRdE_TPD}). 
This explains the 12 orders of magnitude increase in the TPD rate when $m_a$ increases from 1~meV to 1~eV 
and from 1~eV to 1~keV.

\subsubsection{Solar-ALPs at keV mass (non-relativistic)}

Typical helioscope experiments~\cite{CAST-explot} 
can place strong bounds on UR solar-ALPs, in particular for $\ma {\lesssim} 10^{-2}\,\textrm{eV}$. 
The constraints become increasingly weaker in 
the NR regime as $\ma$ increases,
when the coherence wavelength becomes comparable to the detector dimensions.
However, ALPs of $\ma {\sim} {\rm keV}$ can still be produced in the Sun. 
Data from experiments on direct DM searches, as we will show in this work, 
can provide competitive limits from solar-ALPs and complement the ones from helioscopes.

\begin{figure}
	\textbf{(a)}\\
	\includegraphics[width=8.2cm]{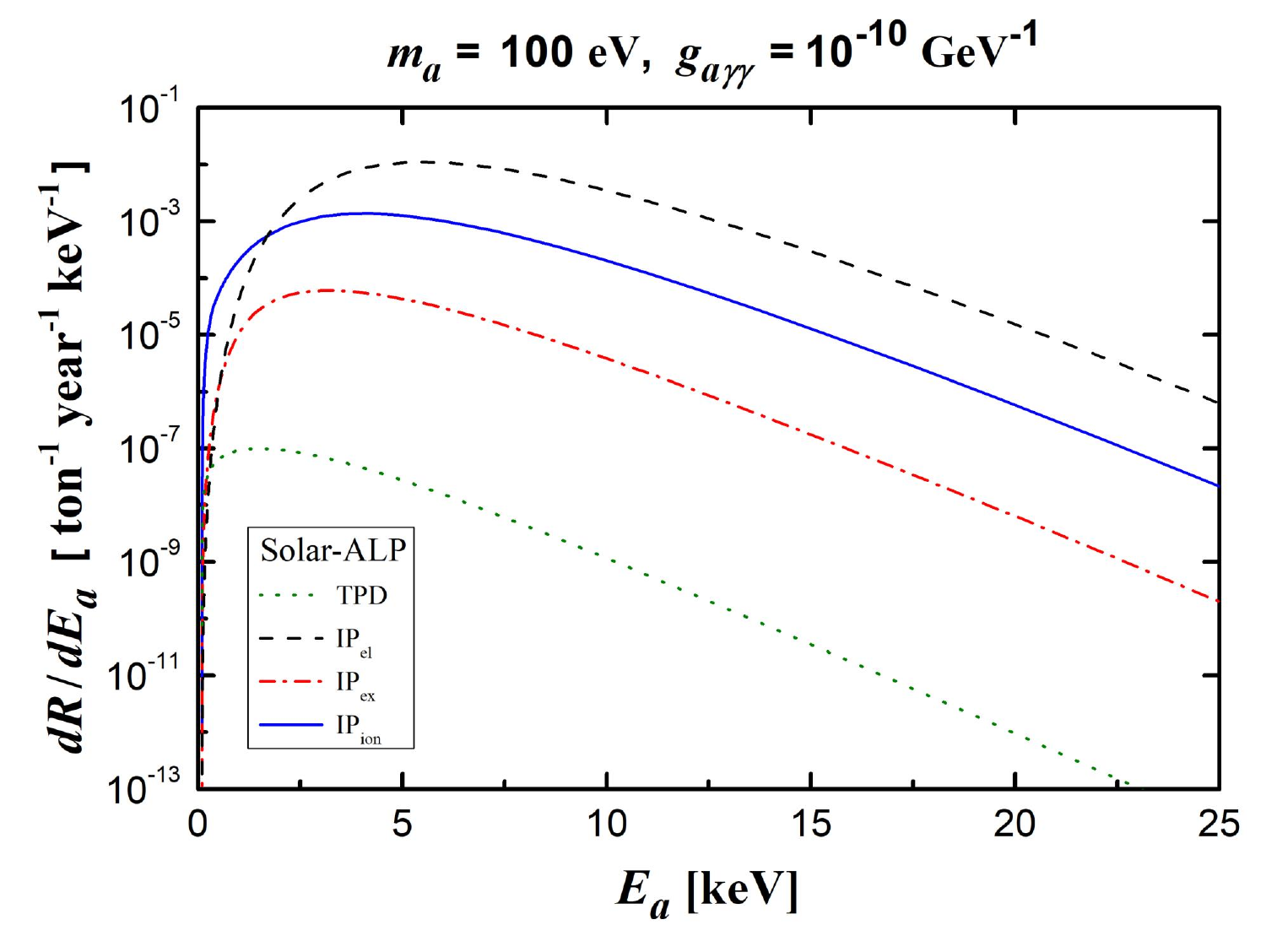}\\
	\textbf{(b)}\\
	\includegraphics[width=8.2cm]{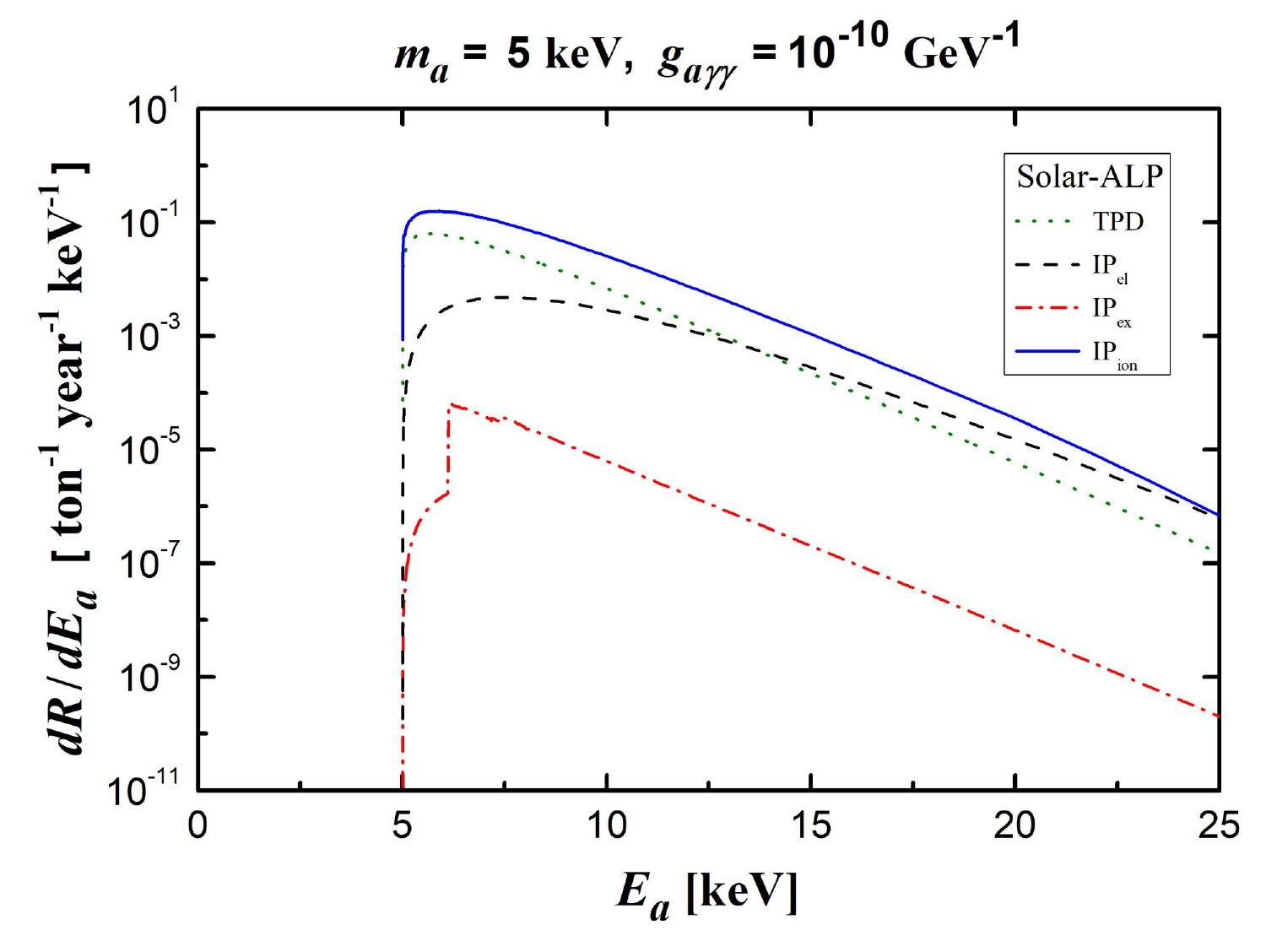}\\
	\caption{ Differential event rates per ton-year of exposure
		for solar-ALPs with the four detection
		channels in liquid xenon, at $\gagg{=}10^{\mbox{-}10}~{\rm GeV^{\mbox{-}1}}$
		with (a) $\ma{=}100~{\rm eV}$ and (b) $\ma{=}5\,{\rm keV}$.
	}
	\label{fig:dR/dT_solar}
\end{figure}

Analytical studies of solar-ALPs cover the entire kinematic regime from NR to relativistic. 
As a result, the derivations of cross sections and rates is involved and
require the incorporation of both atomic longitudinal and transverse responses. 
The case of $\gagg {=} 10^{\mbox{-}10}\,\textrm{GeV}^{\mbox{-}1}$ 
and $\ma {=} 5~{\rm keV}$ is considered in Fig.~\ref{fig:dR/dT_solar}(b). 
The most interesting feature is that $\IPion$
dominates over $\IPel$ in the entire energy range, due to the NR enhancement from the transverse response. 
In addition, since the time dilation effect is not as severe as 
for the case of sub-keV $\ma$, 
the TPD rate is higher than that of $\IPel$.
The $\IPex$ rate remains the least significant. 
The edge structure of $\IPex$ at $\Ea {=} 6.13\,\textrm{keV}$ 
is due to the photon double pole when a $2p$-shell electron is excited to a $5d$ or $6s$ orbital, 
with excitation energy $\sim$4.8~keV. 
The atomic discrete excitations other than the 
electric-dipole transitions are highly suppressed.

The total event rates for the four detection channels 
due to NP solar-ALPs at $\ma {=} {\rm 5 ~  and ~ 10 ~  keV}$ are given
in Table~\ref{tab::Event-Rates}. 
The $\IPion$ channel dominates the event rates for keV-scale solar-ALPs. 
The rates would be dramatically suppressed at $\ma {>} 10 ~ {\rm keV}$
due to the sharp decrease of the solar-ALP fluxes.



\input{iplong-Table1.tex}



\subsection{DM-ALPs}

Under the scenario of DM-ALP~\cite{PRESKILL1983127,ABBOTT1983133,DINE1983137},
experimental searches for galactic DM can place constraints on the ALP parameter 
space~\citep{Graham:2015ouw,Irastorza:2018dyq,Sikivie:2020zpn,Choi:2020rgn,ParticleDataGroup:2022pth}.
IP interactions contribute in probing $\gagg$ in direct DM search experiments.

We first derive the event rates and sensitivity regions under the assumption of DM-ALPs
and independent of other constraints.
We would then study the implications and results 
in the presence of the DM-ALP cosmological stability bound 
in Sect.~\ref{sect::explot-DM_ALP},
and conclude what physics information can be obtained from current and future experiments.

The conventional Maxwellian velocity distribution
is adopted for the DM-ALP spectrum:
\begin{equation}
f\left(\vec{v}_{a}\right) =
\dfrac{1}{K}
~ {\rm exp} [ {-\dfrac{|\vec{v}_{a}+\vec{v}_E|^2}{v_0^2}} ] ~
\Theta\left(v_{\textrm{esc}}-|\vec{v}_{a}+\vec{v}_{E}| \right)~,
\label{eq::DM-halo}
\end{equation}
where $\vec{v}_a$ is the ALP velocity with respect to Earth
and the velocity parameters are $v_0=$220~km/s, $v_E=$232~km/s and 
$v_{\textrm{esc}}=$544 km/s, and $K$ is the normalization factor. 
The galactic DM density is taken to be 
$\rho_{\chi} {=} 0.3~\text{GeV}/\text{cm}^3$.
After averaging out the seasonal effect, 
the isotropic differential number density distribution 
with respect to $v_a$ is
\begin{equation}
	\frac{dn_a(v_a)}{dv_a} = \frac{\rho_{\chi}}{\ma} ~ v_a^2\,\int\,d\Omega_a\,f(\vec{v}_a),
	\label{eq::1d_v_dist.}
\end{equation}
which leads to the DM differential energy spectra: 
\begin{equation}
	\frac{d\phi_a}{d \Ea} = \frac{1}{\ma}  ~ \frac{dn(v_a)}{dv_a}\,.
	\label{eq::dphidEa_ALP}
\end{equation}

Evaluation of  
the differential event rates for the three IP and TPD processes follow Eq.~(\ref{eq:dR/dEa}) 
and Eq.~(\ref{eq::dRdE_TPD}), respectively. 
DM-ALPs are NR or, equivalently, $\gamma {\approx} 1$. The TPD rate is therefore not suppressed by time dilation. 
Since the DM-ALP number density is inversely proportional to $\ma$, the TPD rate 
scales only as $\ma^2$ -- different from $\ma^4$ for solar-ALPs. 
In addition, the flux has no dependence on $\gagg$. The event rates therefore scale as $\gagg^2$.


\begin{figure}

	\includegraphics[width=8.2cm]{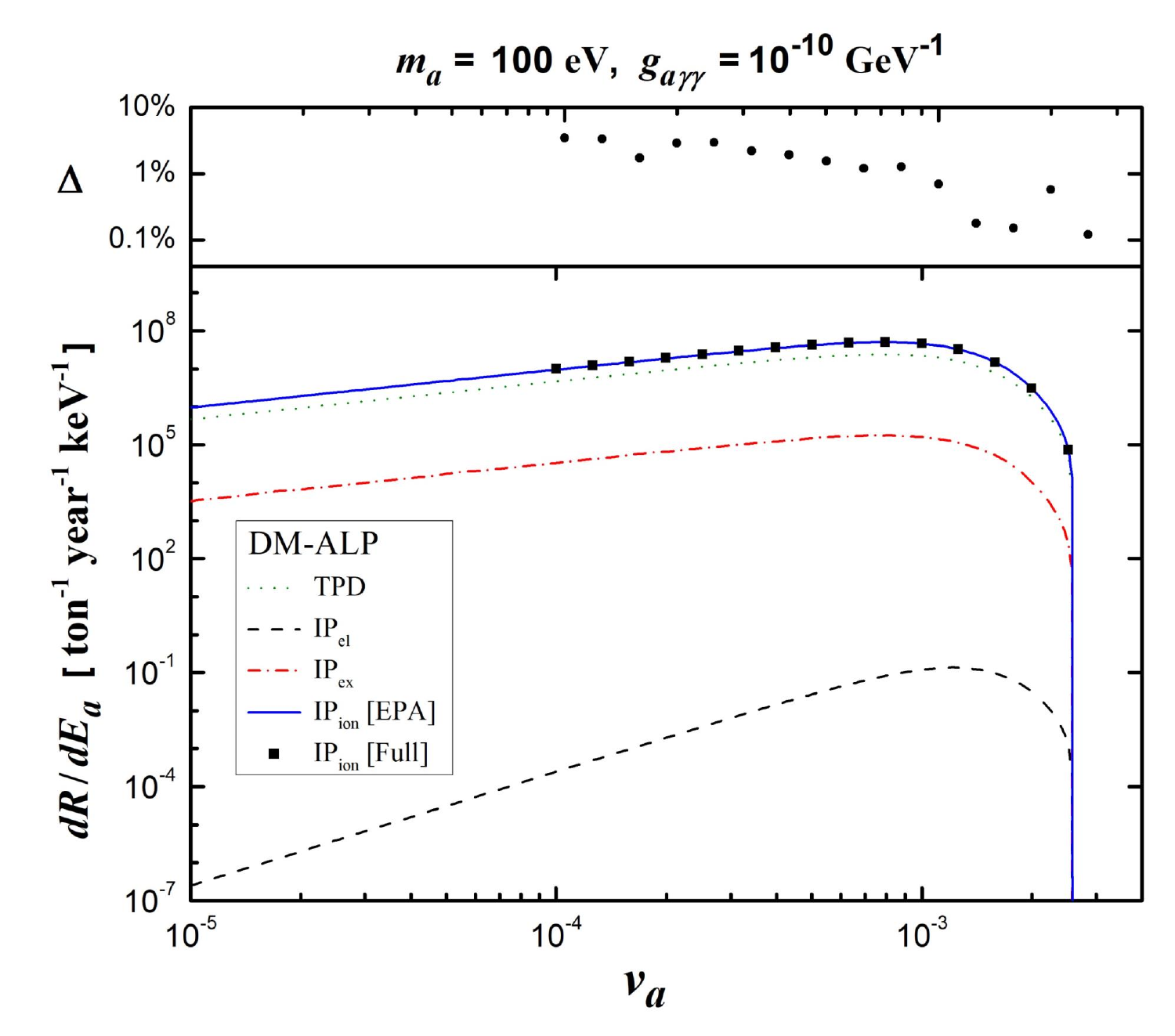}
	\caption{ Differential event rates per ton-year of exposure
		for DM-ALPs with the four detection
		channels in liquid xenon, at
		$\gagg{=}10^{\mbox{-}10}~{\rm GeV^{\mbox{-}1}}$
		and $\ma{=}100~{\rm eV}$ as illustration. 
The $x$-axis is in $v_a {\simeq} \sqrt{2 ( \Ea {-} \ma ) / \ma }$, 
which is suitable for NR kinematics.
The fractional deviation $\Delta$ between the Full and
		EPA calculations of $\IPion$ is shown at the upper panel.
	}
	\label{fig:dR/dT_DM}
\end{figure}

Since DM-ALPs are NR, 
it can be expected that transverse responses in the IP processes are important 
due to the double pole of the photon propagator. 
The differential event rates for the four channels
expected at a liquid xenon detector 
with a ton-year exposure at $\gagg {=} 10^{\mbox{-}10}\,\textrm{GeV}^{\mbox{-}1}$ 
and $\ma {=} 0.1~{\rm keV}$ 
(this set of values for $\ma$ and $\gagg$ satisfies the stability requirement) 
are depicted in Fig.~\ref{fig:dR/dT_DM}.
For clarity, the $x$-axis is converted to the DM velocity 
$\va {\simeq} \sqrt{2 ( \Ea {-} \ma ) / \ma}$ 
with a cut-off at the maximum $v_\textrm{esc}+v_E$. 
Among the three IP channels, $\IPion$ dominates the event rates, while $\IPex$ is sub-leading. 
Both are much larger than $\IPel$ in which there is no enhancement
since the elastic scattering is purely space-like.
In the NR velocity range, the validity of EPA is justified against the full calculations of selected points. 
The percentage deviations, shown in the upper panel, are less than 3\% 
-- consistent with the discussion in 
Sect.~\ref{sec:IID} and Fig.~\ref{fig:vsEPA}.

We argue in Sect.~\ref{subsec:EPA} that, as long as the EPA is a valid approximation, 
the $\IPion$ event rate for NR ALPs in a detector will be twice that of TPD. 
This is verified in Fig.~~\ref{fig:dR/dT_DM}, which implies TPD is 
as important as $\IPion$ to look for DM-ALPs with a nonzero $\gagg$ coupling. 
The efficiency aspects of experimentally tagging two photons for the former and one photon plus one electron 
for the latter will be discussed in Sect.~\ref{sect::expt}.

The total event rates for DM-ALPs at $\ma$=1~eV, 1~keV, and 1~MeV
are given in Table~\ref{tab::Event-Rates}. 
At $\ma {=} 1~{\rm eV}$, the DM-ALP cannot induce any atomic transitions. 
Therefore only $\IPel$ is possible. 
However, the rate is very small compared to that due to a solar-ALP of the same mass
because of the limited phase space.
On the other hand, the TPD rate is much bigger than its solar-ALP counterpart because it does not have the extra 
$10^3$ time dilation suppression factor, and the DM-ALP flux is larger than the solar-ALP flux 
by 8 orders of magnitude. 
This explains the 11-order difference in their TPD rates.   

With $\ma$ at 1~keV and 1~MeV, both $\IPex$ and $\IPion$ are possible and $\IPel$ becomes insignificant. 
As discussed above, $\IPion$ and TPD are the most promising channels for direct detectors and their rates 
only differ by a factor of 2. 
The $\ma^2$ dependence of $\IPion$ and TPD rates is clearly observed. 
The rates of $\IPex$ for these two masses suggest some irregularity. 
This is mostly due to the atomic structure where most discrete excitation energies are in the 10~eV to 10~keV range. 
At $\ma {=} 1~{\rm MeV}$, there is no atomic discrete excitation at $\sim$500~keV(=$m_a/2$) to match 
the double pole kinematics, so there is no enhancement for the transverse response. 
As a result, its rate is even smaller than $\IPel$ 
in which the longitudinal response is larger.




\section{Experimental Constraints}
\label{sect::expt}


\begin{figure}
\includegraphics[width=8.2cm]{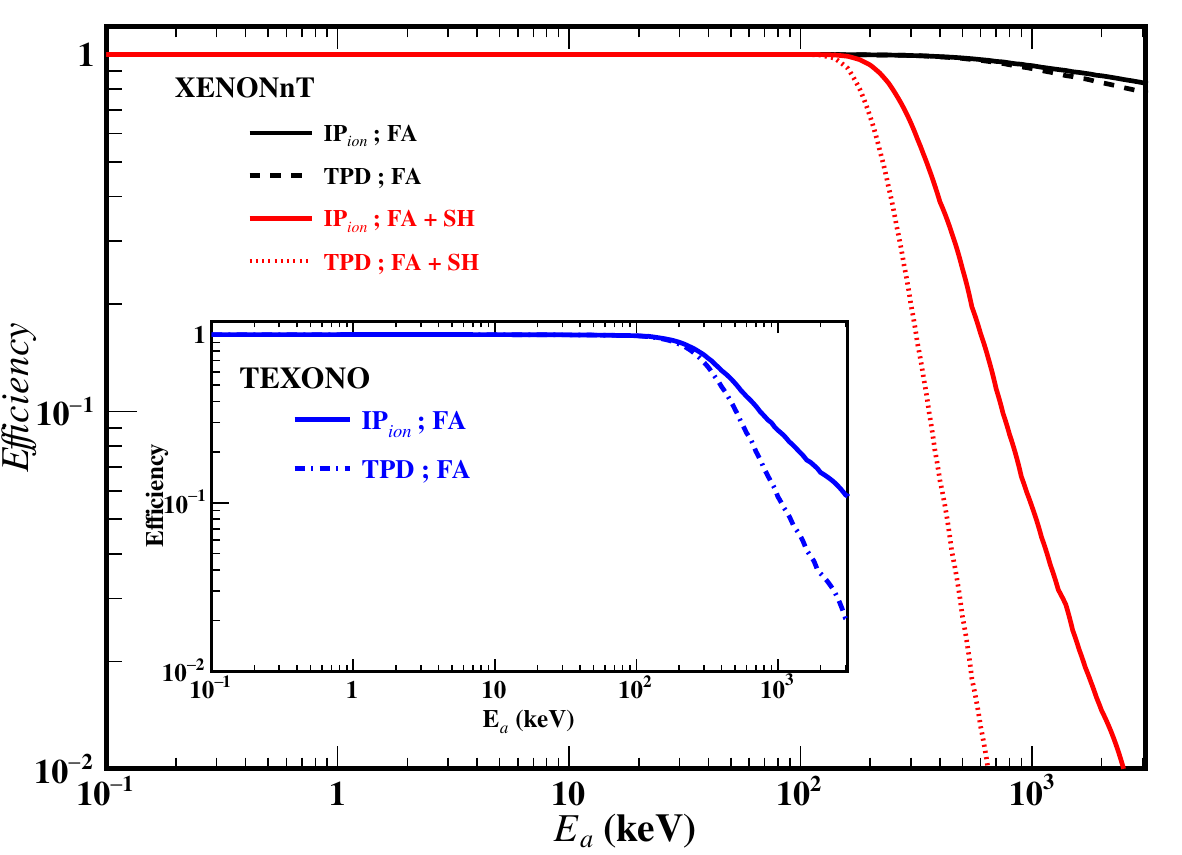}
\caption{
Signal detection efficiency for DM-ALPs for
the leading TPD and $\IPion$ channels
as function of $\Ea$
in XENONnT~\cite{xenon-s2,LXe-spatial-reso} 
and
TEXONO~\cite{texono-pcge, texono-hpge} (inset)
experiments.
Efficiencies due to full absorption (FA)
are applicable to both experiments,
while single-hit (SH) selection applies in addition
to XENONnT.
Signatures for solar-ALPs are below 20~$\keVee$,
and the efficiency is close to unity.
}
\label{fig::Efficiency}
\end{figure}


\subsection{Selected Data}

The following data sets are adopted
for analysis to derive constraints on the $( \ma , \gagg )$ plane:
\begin{enumerate}
\item TEXONO data with (a) point-contact germanium detector at
$300~\eVee {-} 12~\keVee$~\cite{texono-pcge},
and (b) high-purity germanium detector at
12~$\keVee$ to 3000~$\keVee$~\cite{texono-hpge},
selected for having both low threshold and high energy ($\MeVee$) reach
with detectors of excellent energy resolution
(1.98~$\keVee$ at 1~$\MeVee$) for
spectral peak detection.
\item XENONnT data with liquid xenon at
$1 {-} 140~\keVee$~\cite{xenon-s2,LXe-spatial-reso} 
selected for its large exposure while having
low threshold and background.
The background is well-modeled and understood, and
is subtracted for ALP searches.
\end{enumerate}


The energy dependence of signal efficiencies in both  
the TPD and $\IPion$ channels
are depicted in Fig.~\ref{fig::Efficiency}.
For solar-ALPs which are relativistic,
$\IPel$ with a one-photon final state is the dominant channel.
Signatures are continuous distributions with 
the ALP spectra convoluted with
$\IPel$ cross sections.
Signal efficiency for single-hit data selection
is close to unity at this low (${<} 20~\keVee$) energy.
The DM-ALPs are NR and interact pre-dominantly via
the TPD and $\IPion$ channels, 
the signatures of which
are Gaussian peaks at $\ma$ on the total energy depositions
over the continuum background spectra.
Current DM data focuses on ``single-hit'' events
uncorrelated to veto signals from other detector components.
Signal efficiencies of having final-state emissions
with full absorption (FA) in the detectors and tagged as single-hit (SH) events 
have to be evaluated.
These differ among the two channels with their different final-states:
TPD is with both $\gamma$'s each having an energy of $\Ea$/2;
while $\IPion$ has an electron and a photon
each at $\Ea$/2.
In the case of liquid xenon detector, 
this requirement implies that the final states
are fully absorbed within a distance 
corresponding to the spatial resolution 
($\sigma_{x,y} {=} 0.8~{\rm cm}$ and 
$\sigma_z {=} 0.3~{\rm cm}$ at 1~$\MeVee$~\cite{LXe-spatial-reso})
away from the vertices.


\subsection{Solar-ALPs}
 

\begin{figure}
{\bf (a)}\\
\includegraphics[width=8.2cm]{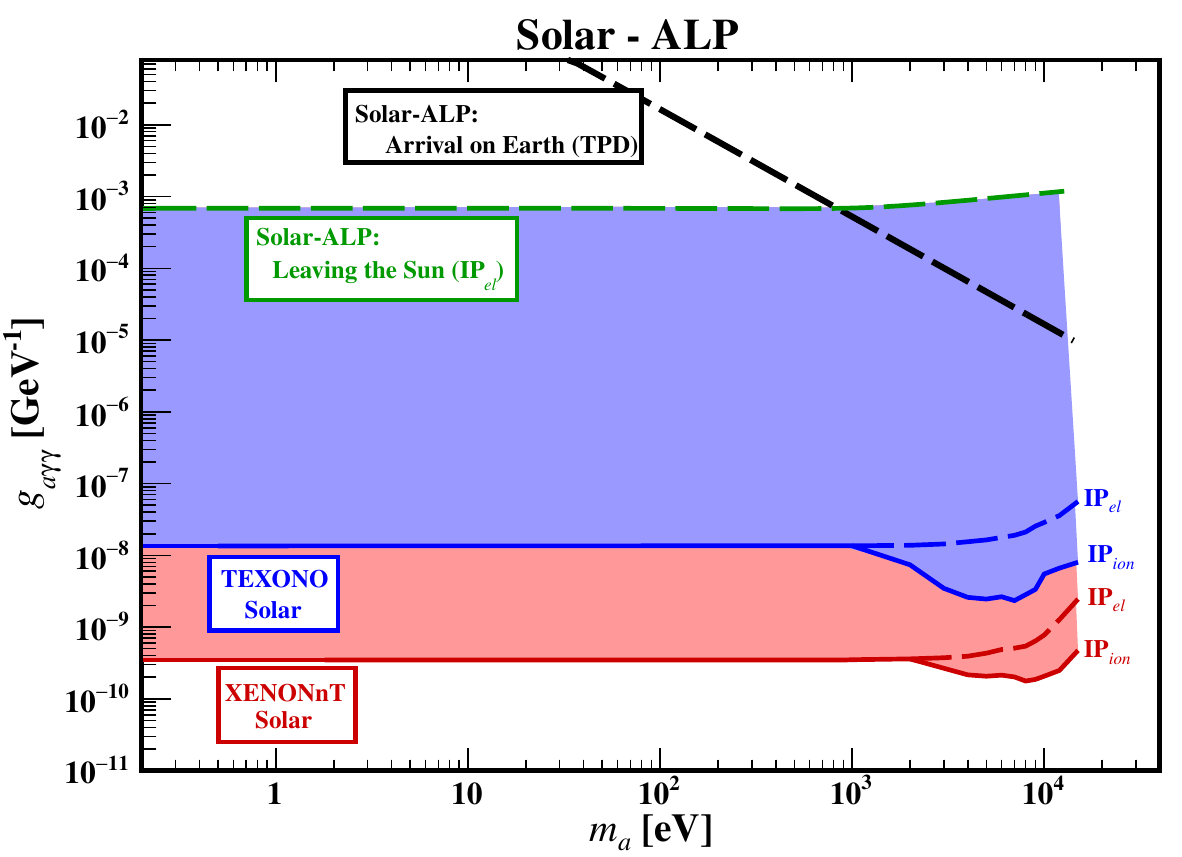}\\
{\bf (b)}
\includegraphics[width=8.2cm]{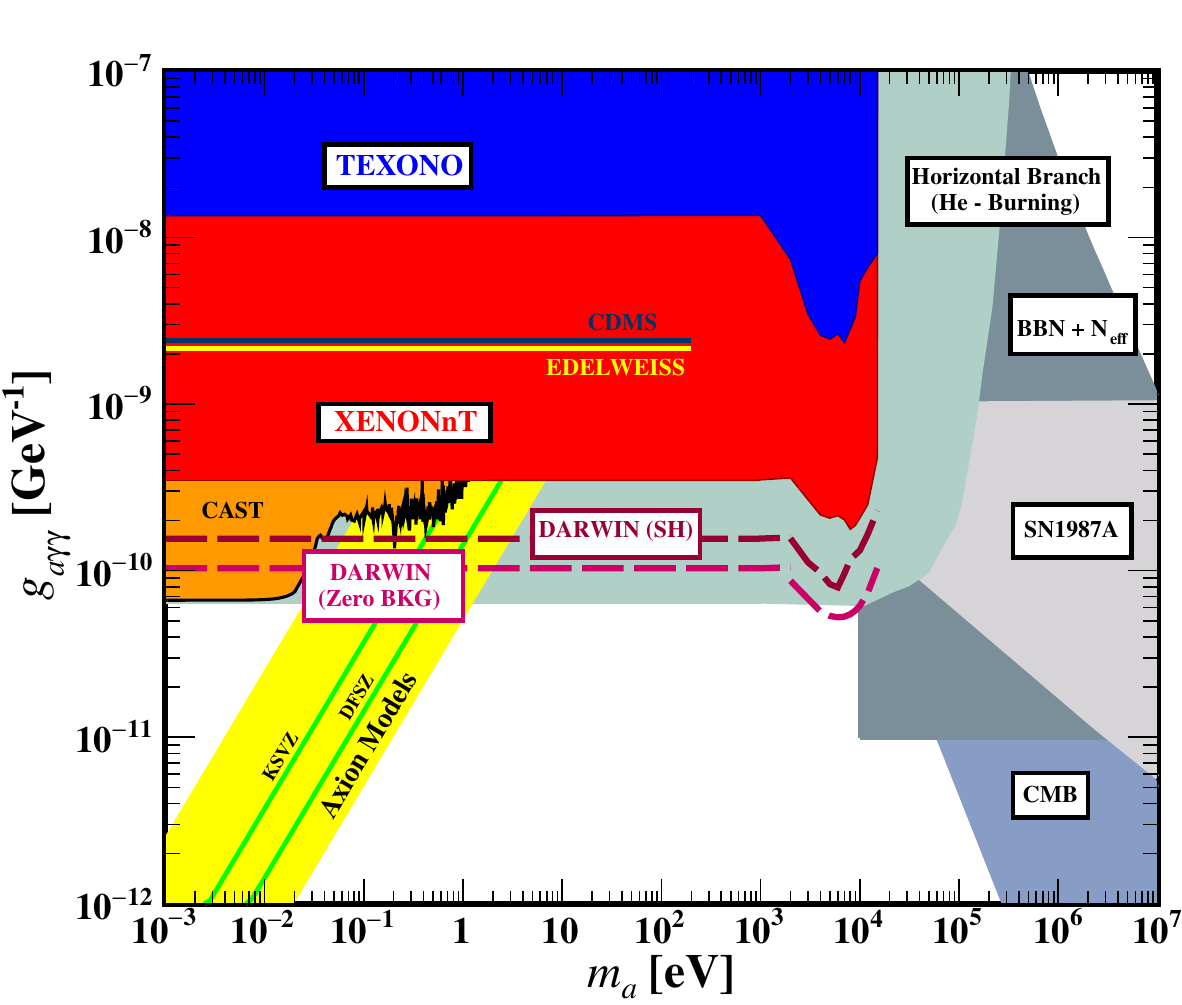}
\caption{
(a)
Standalone sensitivity regions at 90\% CL in $( \ma , \gagg )$ 
from the TEXONO~\cite{texono-pcge, texono-hpge}
and XENONnT~\cite{xenon-s2,LXe-spatial-reso} experiments
with solar-ALPs.
Contributions from the leading channels
of $\IPel$ and $\IPion$ are displayed.
Upper reaches of the sensitivity regions
due to survival from the Sun is shown.
(b)
Exclusion plot in $( \ma , \gagg )$ at 90\% CL,
showing the solar-ALP limits
from TEXONO   
and XENONnT 
experiments.
The astrophysical and cosmological 
bounds~\cite{Irastorza:2018dyq,astro-explot1,Jaeckel:2017tud,Depta:2020wmr}
are the light shaded regions.
The predicted band for QCD axions~\cite{GammaVacuum2}
is in yellow.
Superimposed are the current constraints from solar-ALPs with
Bragg scattering~\cite{CDMS:2009fba,Armengaud:2013rta}
and helioscope~\cite{CAST-explot} experiments,
as well as the sensitivity reaches of
the DARWIN project~\cite{darwin}
at standard SH-selection and
zero-background scenarios.
}
\label{fig::solar-alp-explot}
\end{figure}


The exclusion regions at 90\% confidence level (CL)
from the TEXONO and XENONnT data on 
solar-ALP $\gagg$ are derived.
The signal detection efficiencies are 100\% as noted.
The standalone sensitivity regions are presented 
in Fig.~\ref{fig::solar-alp-explot}a.
The relevant mass range for solar-ALPs
is $\ma {<} 10^4 ~ {\rm eV}$,
where the blue and red lines represent the constraints from
TEXONO and XENONnT, respectively. 
Also displayed are the upper reaches 
of the exclusion sensitivity, 
which are constrained by the solar-ALPs being able to:
(a) leave the solar surface, 
limited by $\IPel$ in the Sun (green dotted line), 
and
(b) arrive on Earth, limited by TPD in space (black dotted line).

The leading sensitivities are from 
the $\IPel$ and $\IPion$ channels
at the relativistic ($\ma {<} 1~{\rm keV}$)
and NR ($\ma {>} 1~{\rm keV}$) ranges, respectively. 
Limits derived from XENONnT far exceed that of TEXONO
over the entire solar-ALP $\ma$-range, due to its large exposure
and lower background.

The solar-ALP constraints in the global context are illustrated
in Fig.~\ref{fig::solar-alp-explot}b,
together with
astrophysical and cosmological
bounds~\cite{Irastorza:2018dyq,astro-explot1,Jaeckel:2017tud,Depta:2020wmr}
as well as predictions from QCD-axion models~\cite{GammaVacuum2}. 
The $\IPel$ and $\IPion$ channels with solar-ALPs
significantly improves on $\gagg$ over
the Bragg-scattering constraints from CDMS~\cite{CDMS:2009fba}
and EDELWEISS~\cite{Armengaud:2013rta}.
It also extends the detectable window in $\ma$
from $1~{\rm eV}$ to $\mathcal{O}$(10~keV)
beyond the reach of
the CAST helioscope experiment~\cite{CAST-explot}.


\subsection{DM-ALPs}
\label{sect::explot-DM_ALP}


\begin{figure}
{\bf (a)}\\
\includegraphics[width=8.2cm]{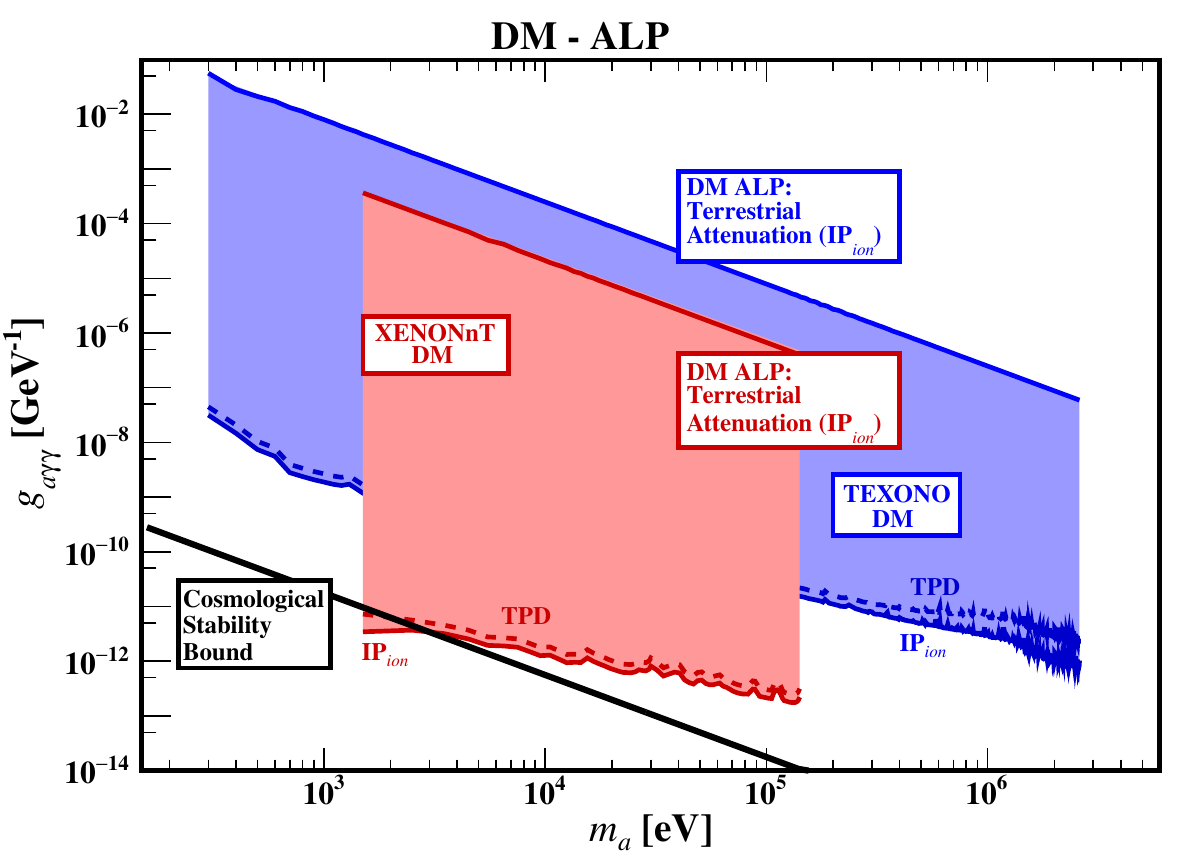}\\
{\bf (b)}
\includegraphics[width=8.2cm]{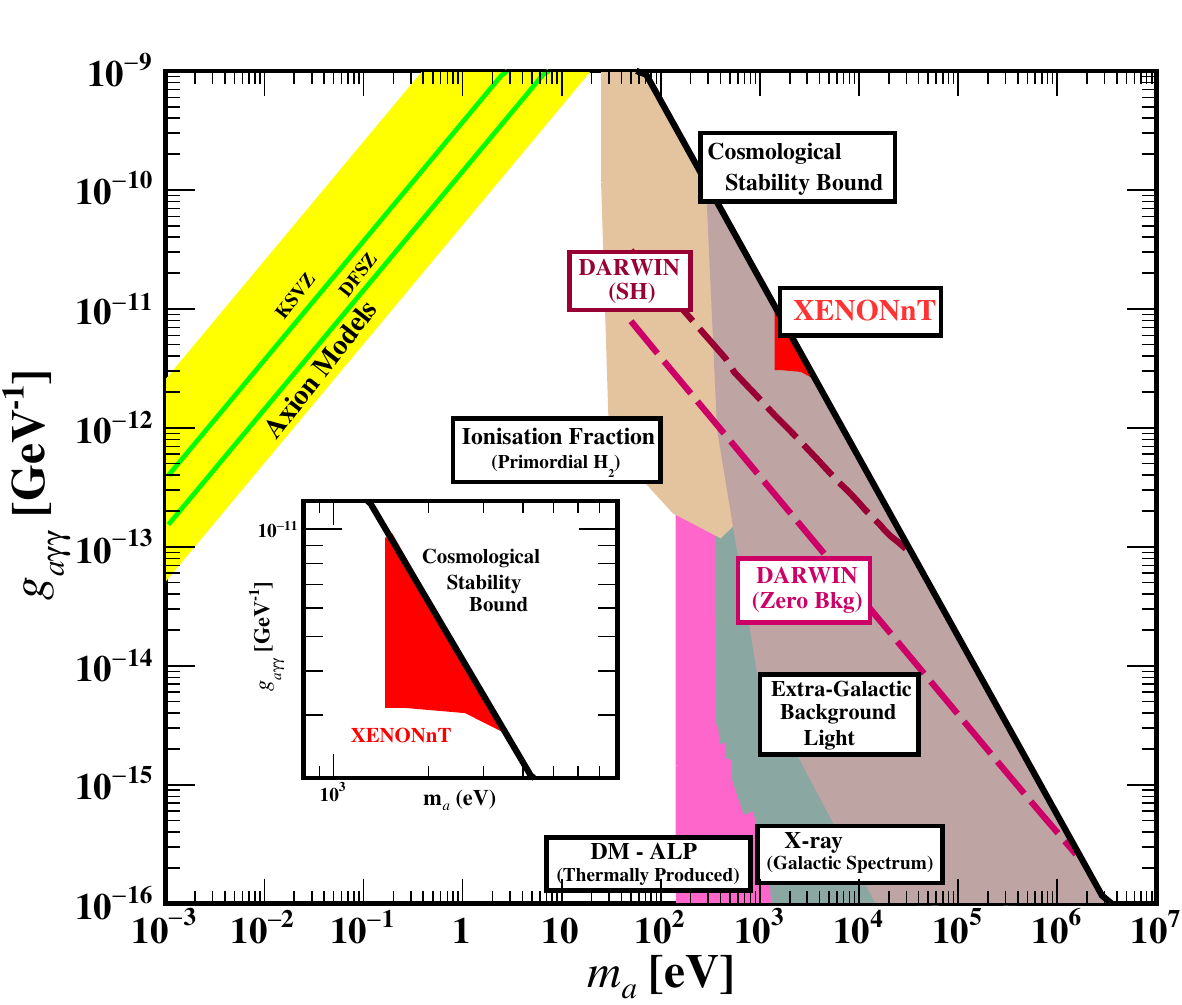}
\caption{
(a)
Standalone sensitivity regions at 90\% CL in $( \ma , \gagg )$
from the TEXONO~\cite{texono-pcge, texono-hpge}
and XENONnT~\cite{xenon-s2,LXe-spatial-reso} experiments
with DM-ALPs.
Contributions from the leading channels
of $\IPion$ and TPD are displayed.
Upper reaches of the sensitivity regions
due to survival of terrestrial attenuation effects
are shown.
The cosmological
stability bound~\cite{Masso:1995tw,astro-explot1,Arias:2012az}
is denoted by the bold black line.
(b)
Exclusion plot in $( \ma , \gagg )$ at 90\% CL which
results from the scenario of DM-ALPs.
The stability bound (black) dictates that
only a small region (red) is excluded by 
XENONnT~\cite{xenon-s2,LXe-spatial-reso}.
The astrophysical and cosmological 
constraints~\cite{Irastorza:2018dyq,Arias:2012az}
which are consequences of DM-ALPs are included. 
The predictions from QCD axions~\cite{GammaVacuum2}
are displayed as the yellow band.
Superimposed are the sensitivity reaches of
the DARWIN project~\cite{darwin}
at standard SH-selection and zero-background scenarios,
indicating that a new detection window is opened and
substantial region can be probed.
}
\label{fig::dm-alp-explot}
\end{figure}


\begin{figure}
\includegraphics[width=8.2cm]{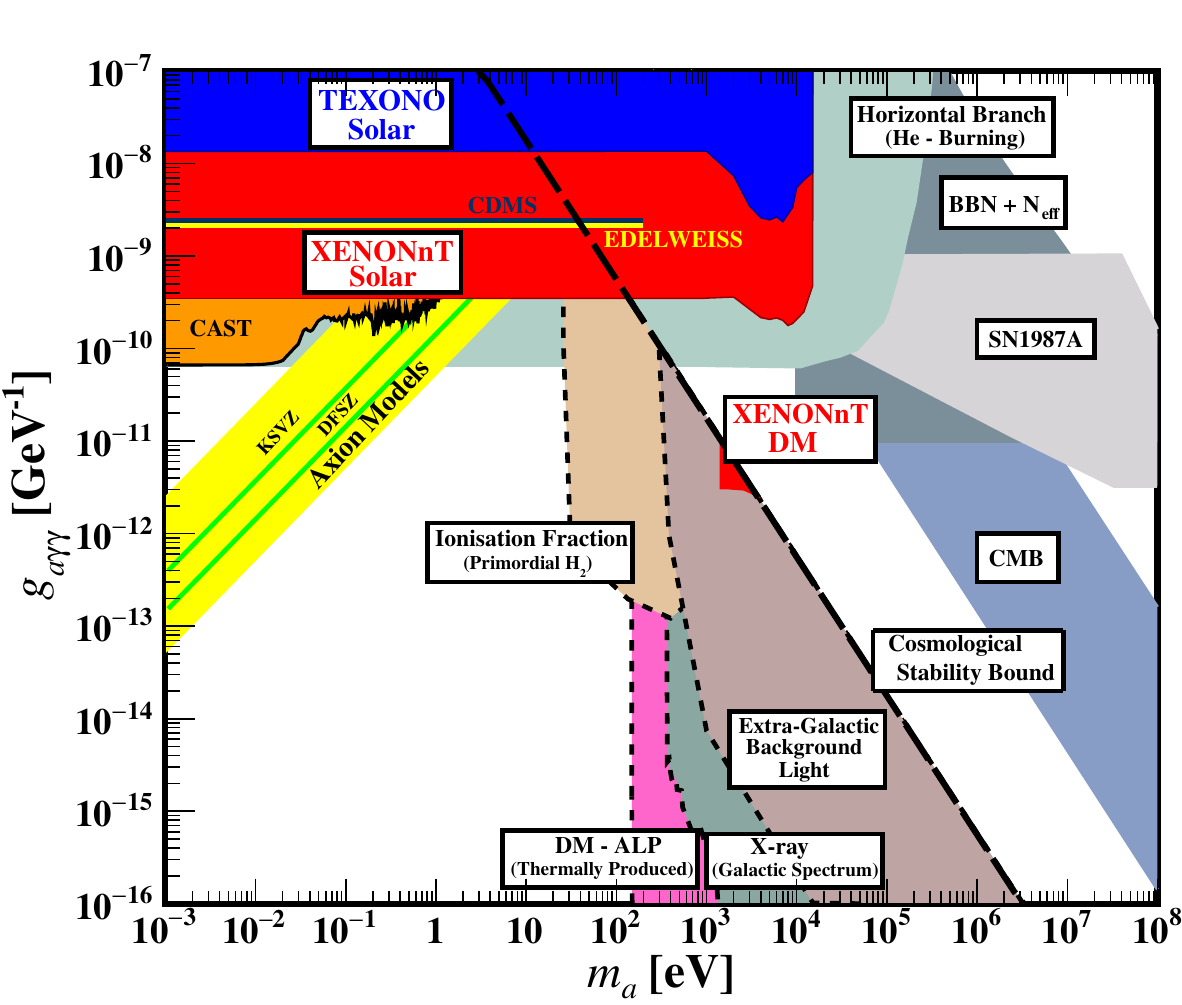}
\caption{
Summary plot of existing constraints on $\gagg$ versus $\ma$ in ALPs.
Those in dotted lines only apply under the assumption that ALPs are the
cosmological DM~\cite{Arias:2012az}.
The yellow band are predictions from QCD axion models~\cite{GammaVacuum2}.  
}
\label{fig::combined_plot}
\end{figure}


With the detection efficiencies of Fig.~\ref{fig::Efficiency}
taken into account, 
the standalone sensitivity regions for DM-ALPs analysis
independent of other processes are derived. 
The probed parameter space at 90\% CL
of the leading channels TPD and $\IPion$ for DM-ALPs 
by the TEXONO and XENONnT data 
are presented by the blue and red regions, respectively,
in Fig.~\ref{fig::dm-alp-explot}a.
The relevant mass range for DM-ALPs is
$ 10^4~{\rm eV} {\leq} \ma  {\leq} 10^7~{\rm eV}$,
corresponding to the measurement dynamic range of the experiments.
The upper reaches of the sensitivity regions are bounded
by terrestrial attenuation by the Earth and its atmosphere
before the DM-ALPs can reach the detectors.
XENONnT provides better sensitivities,
at $\ma {\sim} 1 {-} 100 ~ {\rm keV}$.
TEXONO, being a surface experiment with larger dynamic range, 
is sensitive to higher $\gagg$ and
can explore an extended mass range of 
$\ma {\sim} 300 ~ {\rm eV} {-} 3 ~ {\rm MeV}$.

However, it is necessary to place this projected standalone sensitivity 
in relation to other astrophysical and cosmological constraints.
The decay of ALPs with $\ma {<} 1 ~ {\rm MeV}$ 
to $e^+ e^-$ via $\gaee$ is not allowed by kinematics,
but ALPs of any mass can have TPD via $\gagg$.
This places severe requirement on ALP stability
$-$ the TPD lifetime ($ 1 / \Ga2g$) has to be longer than
the age of the Universe~\cite{Masso:1995tw,astro-explot1,Arias:2012az}
in order for the DM-ALP
to reach and be observable in terrestrial experiments.
This implies part of the ($m_a , \gagg$) parameter space is not
accessible by the direct experimental searches of DM-ALPs.
As an illustration, the case of
$m_a {=} 1 ~ \textrm{keV}$, 
$\gagg {\gtrsim} 2 \times 10^{\mbox{-}11} \textrm{GeV}^{\mbox{-}1}$ 
is not viable.
The DM-ALPs would not reach the detectors and no physics information 
can be extracted.\footnote{The relatively-long-lived ALPs can be studied by
other means such as in colliders and from probes of the early Universe.}

A summary of experimental constraints on $\gagg$ from DM-ALPs
together with  predictions from QCD-axion models~\cite{GammaVacuum2}
are presented in Fig.~\ref{fig::dm-alp-explot}b. 
The TPD stability bound is marked as the bold black line. 
The $\IPion$ channel 
opens a large unexplored detection window for DM-ALPs.
The current sensitivities, however, do not yet match 
those required by the stability bound,
and no physics constraints can be placed.
A notable crack is at a small corner at $\ma {\sim} 10^{3}~{\rm eV}$,
where the sensitivity from XENONnT data~\cite{xenon-s2,LXe-spatial-reso} exceeds that of stability bound,
so that the parameter space in red is probed and excluded.
Other astrophysical and cosmological bounds 
under the DM-ALP scenario~\cite{Irastorza:2018dyq,Arias:2012az}
are also included in Fig.~\ref{fig::dm-alp-explot}b. 
The constraints are valid only at regions where
$\gagg$ is weak enough to survive the cosmological stability bound.


\section{Summary and Prospects}

In this work,
we explored new detection channels 
to probe $\gagg$ with terrestrial experiments,
opening up a
large parameter space in $( \ma , \gagg )$ 
for laboratory-based experimental searches 
of solar-ALPs and DM-ALPs.
Both results are combined into the summary plot 
of Fig.~\ref{fig::combined_plot}.
The constraints in dotted lines represent those which
only apply under the assumption that ALPs are the
cosmological DM and 
are stable relative to the age of the Universe.
Though the astrophysical and cosmological constraints are in general
more stringent than the laboratory limits, 
they usually have strong model and parameter dependence.
For instance, the ``DM-ALP (Thermally Produced)'' constraint 
assumes thermal production of DM-ALP in the early universe
such that its density cannot exceed that implied by
the DM relic abundance~\cite{astro-explot1}.
This bound would be completely evaded under 
DM-ALP production scenarios with non-thermal processes~\cite{Arias:2012az}.

Experimentally,
the $\IPion$ channel with NR ALPs 
offers a very distinct signature:
an electron and a photon with equal energy
originated from a common vertex.
This feature can be used for further background suppression
while retaining good signal efficiency.
The projected sensitivities 
of the next generation liquid xenon project 
DARWIN~\cite{darwin} at 200~ton-year exposure
and 50~$\eVee$ threshold  
are superimposed in 
Figs.~\ref{fig::solar-alp-explot}b\&\ref{fig::dm-alp-explot}b,
showing one with typical single-hit selection 
with the projected background subtracted,
and another with an idealized zero-background 
all-multiplicity full-efficiency measurement.
In particular, Fig.~\ref{fig::dm-alp-explot}b indicates
that the $\IPion$ sensitivities of future projects
would exceed the cosmological stability bound for DM-ALPs.
A new detection window would be opened by the IP processes, 
and an unexplored parameter space can be studied.

Recent works~\cite{gaeeloop1,gaeeloop2} identify that 
$\gaee$ coupling at the quantum one-loop level
can produce experimental signatures which resemble those
from $\gagg$ interactions. 
Accordingly, experimental constraints 
such as those from IP effects of this work
can be translated to bounds on $\gaee$.
This will be one of the themes 
for our future research efforts.

\section{Acknowledgement}

This work was supported in part under Contracts 
106-2923-M-001-006-MY5, 110-2112-M-001-029-MY3,
108-2112-002-003-MY3, 109-2112-M-259-001 and 110-2112-M-259-001
from the Ministry of Science and Technology, 
2019-20/ECP-2 and 2021-22/TG2.1 
from the National Center for Theoretical Sciences, 
and the Kenda Foundation (J.-W. C.) of Taiwan; 
Contract F.30-584/2021(BSR), UGC-BSR Research Start Up Grant, India (L.S.);
and 
the Canada First Research Excellence Fund 
through the Arthur B. McDonald Canadian Astroparticle Physics 
Research Institute (C.-P. W.).

\bibliography{iplong_1,iplong_2}

\end{document}

%% file: iplong-Table1.tex

\begin{table}
\centering
\caption{
Event rates of the four detection channels in liquid xenon
for solar- and DM-ALPs,
at $\gagg {=} 10^{\mbox{-}10}~{\rm GeV}^{\mbox{-}1}$.
For $\IPex$, only the dominant final states by
electric dipole excitations are included.
}
\begin{center}
\begin{tabular}{|lcccc|}
\hline
& \multicolumn{4}{c|}{\rm Event Rates ($ {\rm ton^{\mbox{-}1} year^{\mbox{-}1} }$)  } \\
& \multicolumn{4}{c|}{\rm Detection Channels} \\
$\ma$ & ~~ TPD ~~ & ~~ $\IPel$  ~~ & ~~ $\IPex$  ~~ & ~~ $\IPion$  ~~ \\
\hline \hline
\multicolumn{5}{|c|}{\bf solar-ALP} \\
1~meV  & $\mathcal{O}(10^{\mbox{-}27})$
& $6.6 {\times} 10^{\mbox{-}2}$ & $2.9 {\times} 10^{\mbox{-}4}$ & $7.8 {\times} 10^{\mbox{-}3}$ \\
1~eV &  $\mathcal{O}(10^{\mbox{-}15})$
& $6.6 {\times} 10^{\mbox{-}2}$ & $2.9 {\times} 10^{\mbox{-}4}$ & $7.8 {\times} 10^{\mbox{-}3}$ \\
1~keV & $2.8 {\times} 10^{\mbox{-}3}$
& $6.4 {\times} 10^{\mbox{-}2}$ & $2.6 {\times} 10^{\mbox{-}4}$ & $1.3 {\times} 10^{\mbox{-}2}$ \\ 
{5~keV} & $1.8 {\times} 10^{\mbox{-}1}$ 
& $2.4 {\times} 10^{\mbox{-}2}$ & $9.4 {\times} 10^{\mbox{-}5}$ & $5.1 {\times} 10^{\mbox{-}1}$ \\
{10~keV} & $8.9 {\times} 10^{\mbox{-}2}$ 
& $1.9 {\times} 10^{\mbox{-}3}$ & $7.8 {\times} 10^{\mbox{-}5}$ & $3.8 {\times} 10^{\mbox{-}1}$ \\ \hline
\multicolumn{5}{|c|}{\bf DM-ALP} \\
1~eV & $2.5 {\times} 10^{\mbox{-}4}$ & $\mathcal{O}( 10^{\mbox{-}14} )$ & 0  & 0 \\
1~keV & 250 & $1.7 {\times} 10^{\mbox{-}5}$ & $1.6 {\times} 10^{\mbox{-}3}$ &  500  \\
1~MeV & ~ $2.5 {\times} 10^{8}$ ~ &  ~ $9.9 {\times} 10^{\mbox{-}5}$ ~
& ~ $2.2 {\times} 10^{\mbox{-}10}$ ~ & ~ $5.0 {\times} 10^{8}$ ~ \\ \hline
\end{tabular}
\end{center}
\label{tab::Event-Rates}
\end{table}
